\documentclass[]{elsarticle}

\usepackage{epsfig}
\usepackage{graphicx}
\usepackage{amsfonts,amsmath,amssymb,amsthm}
\usepackage{mwe}
\usepackage[export]{adjustbox}
\usepackage{geometry}
\usepackage{booktabs}
\usepackage{enumitem}
\usepackage{parskip}
\usepackage[pagewise,left,displaymath]{lineno}
\modulolinenumbers[1]
\usepackage[table]{xcolor}

\newtheorem{Prop}{Proposition}
\usepackage[ruled]{algorithm2e}
\usepackage{algorithmic}

\DeclareMathOperator*{\argmax}{argmax}

\definecolor{mycolor1}{cmyk}{0, 0.0808, 0.1429, 0.1412}
\definecolor{mycolor2}{cmyk}{0, 0.1808, 0.1429, 0.1412}
\definecolor{mycolor3}{cmyk}{0, 0.0808, 0.0429, 0.0412}
\journal{Information Sciences}

\newcommand\boldred[1]{\textcolor{black}{}}
\newcommand\boldblack[1]{\textcolor{black}{}}
\newtheorem{thm}{Theorem}[section]

\newtheorem{lem}[thm]{Lemma}

\newcommand{\ie}{\emph{i.e. }}								  
\newcommand{\eg}{\emph{e.g. }}								  

\newcommand{\real}{\ensuremath{\mathbb{R}}}						 

\newcommand{\Log}[2]{\textnormal{Log}_{#1}(#2)}			   
\newcommand{\Exp}[2]{\textnormal{Exp}_{#1}(#2)}			   
\newcommand{\Sphere}{\mathcal{S}}							  
\newcommand{\PP}{\mathcal{P}}						   
\newcommand{\HH}{\mathcal{H}}						   

\begin{document}
	
	\begin{frontmatter}
		\title{Bayesian Regression and Classification Using
			Gaussian Process Priors  Indexed by   Probability Density Functions}
		
		
		\author[mymainaddress]{A. Fradi}
		\author[mymainaddress]{Y. Feunteun}
		\author[mymainaddress]{C. Samir}
		\author[mymainaddress]{M. Baklouti}
		\author[mysecondaryaddress]{F. Bachoc}
		\author[mysecondaryaddress]{J-M Loubes}

		\address[mymainaddress]{CNRS-LIMOS, UCA, France}
		\address[mysecondaryaddress]{Institut de
			Mathématiques de Toulouse, France}

		\begin{abstract}
			In this paper, we  introduce the notion of   Gaussian processes indexed  by probability density functions  for extending the  Mat\'ern family of covariance functions. We use  some tools from  information geometry to improve the  efficiency and the computational aspects of the Bayesian learning model. We particularly show how a  Bayesian inference with a  Gaussian process prior  (covariance parameters estimation  and prediction) can be put into action on the  space of probability density functions. Our framework has the capacity of classifiying and infering on  data observations that lie on nonlinear subspaces.
            Extensive experiments on multiple synthetic, semi-synthetic and real data demonstrate the effectiveness and the efficiency of the proposed methods in comparison with current state-of-the-art methods.
            
		\end{abstract}
		
		
	\end{frontmatter}

\section{Introduction}
In recent years,  Gaussian processes on manifolds have become very
popular in various fields including machine learning, data mining, medical imaging,  computer
vision, etc. The main purpose consists in inferring the
unknown target value  at an observed location on the manifold as a prediction by conditioning on known inputs and targets. 
The  random field, usually
Gaussian, and the forecast can be seen as the posterior mean, leading to an optimal  unbiased predictor~\cite{bachoc2017gaussian,AbtWel1998}. Bayesian regression and classification models
 focus on the use of priors for the parameters to define  and estimate a conditional predictive expectation. In this work, we  consider a very common  problem in  several contexts of applications in science and technology: learning  a Bayesian regression and classification models with Probability Density Functions as inputs.

Probability Density Functions (PDFs) are inherently infinite-dimensional objects. Hence, it is not straightforward  to extend traditional  machine learning methods from finite vectors to functions. For example, in functional data analysis~\cite{Anuj-book-2016} with applications in medical~\cite{Samir-wacv-16,Ramsay-1991}, it is very common to compare/classify functions. The mathematical formulation leads to a wide range of applications where it is crucial to characterize a population or to build predictive models. 
In particular, multiple frameworks exist for comparing PDFs in different representations
 including Frobenius, Fisher-Rao, log-Euclidean, Jensen-Shannon
and Wasserstein distances~\cite{Srivastava-2007,Mitsuhiro-Geodesic-2015,Samir-wacv-16,Nguyen-Div-JS,bachoc2017gaussian}. In this work, we extend this formulation to  PDFs space $\PP$ with the  Mat\'ern  covariance functions.

\textcolor{black}{There is a rich literature on statistical inference on manifolds among which the  Fisher information matrix~\cite{Rao-45} has played a central role. Recently, there has been  increasing interest in applying information geometry for machine learning and data mining tasks~\cite{Amari-Rao-87,Nihat-book,Amari-book,Barbaresco-2013}. The Fisher information matrix determines a Riemannian structure on a parametrized space 
 of probability measures. Study of geometry of $\PP$
 with the Riemannian structure, which we call information geometry, contributes greatly to statistical inference, refer to~\cite{Nihat-book,Amari-book} for more details. Such methods are based on parametric models  that are of great interest in many applications. However,  aspects of PDFs  other than parametric families  may be important in various  contexts~\cite{Nguyen-Div-JS,Nielsen2013,Srivastava-2007,shishido2005,Samir-wacv-16,Fukumizu-2010}.  In particular,  
 the consistency of regression and classification with PDFs inputs was established in~\cite{DBLP:journals/corr/SutherlandOPS15,DBLP:conf/aistats/OlivaNPSX14,pmlr-v31-poczos13a} with the help of kernel density estimation~\cite{Botev-2010}.
 More recently,~\cite{bachoc2017gaussian} studied the dissimilarity between PDFs  with  the Wasserstein distance and~\cite{Zhang-2019} used a nonparametric framework to compare spherical populations.}

The main aim of this paper is to learn a Bayesian inference on Gaussian processes. For instance,  one can think of a Gaussian process as defining PDFs and inference  taking place directly in the function-space.  Moreover, the index space is that of PDFs when choosing the underlying metric in order to evaluate the dissimilarity between them~\cite{Atkinson-81-RaoDistance}. The only drawback is that performing Kriging on PDFs space $\mathcal{P}$ is not straightforward due to its geometry. \textcolor{black}{For this end, we  exploit an isometric embedding by combinng the square root transform~\cite{Paper-Bhattacharyya43} and the distance induced by the Fisher-Rao metric which make the covariance function non-degenerate and simplify the  optimization process.} 

Gaussian processes (GPs) have been widely used to provide a  probabilistic framework for a large variety of machine learning methods~\cite{rasmussen06gaussian}. Optimization techniques are usually required to fit a GP model $Z$, that is to select a GP covariance function.  For $p_i$ and $p_j$ in $\PP$, the main issue would be to build a proper covariance between $Z(p_i)$ and $Z(p_j)$.  In particular, this covariance can define a notion of  stationarity for the process. Another important task is the classification process where we wish to assign an input PDF $p_i$ to one of the given classes~\cite{NIPS2011_4241}.

To search for the covariance function hyperparameter, we use several  methods for maximizing the marginal likelihood. Our aim is then to select those optimizing  performance criteria for regression and classification: The first method is based on the gradient descent for finding a local maximum of the marginal likelihood. The second method is  a special case of MCMC methods, called Hamiltonian Monte-Carlo (HMC)~\cite{Duane1987216}. The objective is to perform sampling from a probability distribution for
which the marginal likelihood and its gradient  are known. The latter has the advantage to simulate from a physical system governed by Hamiltonian dynamics.

The remainder of the paper is organized as follows. In Section~\ref{sec:Proposed}, we introduce the problem formulation and we give a background of some Riemannian representations. Section~\ref{sec:GP} extends the usual notion of GPs indexed by finite vectors to those indexed by PDFs with theoretical results for the Mat\'{e}rn covariance function.
We also give details of  
the proposed model for predicting and classifying  PDFs as well as estimating the  covariance function parameters. 
In Section~\ref{sec:results}, 
we present and discuss experimental results with some comparison studies. We conclude the paper in Section~\ref{sec:clc}.

\section{Problem formulation and geometry background}
\label{sec:Proposed}
\noindent  Let  $p_1,\dots, p_n$ denote a finite set of  observed  PDFs and $y_1, \dots , y_n$ denote their corresponding outputs with real values (quantitative or qualitative). 
In this work, we focus on \textbf{nonparametric PDFs} restricted to be  defined on $\Omega=[0,1]$. 
Our main goals throughout this paper are: i)  Fitting the proposed model's parameters
in order to
better explain the link between $p_i$ and $y_i$, $i=1,\dots,n$, ii)
evaluating the corresponding predictive expectation at
 an unobserved PDF $p^* \in \mathcal{P}$ and iii) 
studying the properties of the GP with the Mat\'ern covariance function.
In the particular case where $y_i \in \{-1,+1\}$, we will assign each unobserved PDF $p^*$ to its predicted class after learning the model parameters. 
To reach such goal, we follow  the same idea of nonparametric information geometry that has been discovered by~\cite{Rao-45} and  developed later in other works, see for example~\cite{Friedrich91,Srivastava-2007,Mitsuhiro-Geodesic-2015,Nihat-book,Fukumizu-2013}. Thus, the notion of similarity/dissimilarity between any $p_i$ and $p_j$ is measured  using the induced Rao distance~\cite{Rao-82,Atkinson-81-RaoDistance}  between them  on the underlying space. 
In this paper, we look at the space of PDFs as a Riemannian manifold, as detailed in the next section, which plays an important role in the proposed methods.
\subsection{Riemannian structure of  PDFs space}
\label{sec:background}
For more details about the geometric structure concerning the Fisher information metric, refer to~\cite{Friedrich91,shishido2005,Bauer-2016}. For example,~\cite{Friedrich91} showed that  $\PP$ with a Riemannian structure has a positive constant curvature. Furthermore, the action of orientation preserving diffeomorphism  acts by isometry on $\PP$   with respect to the Fisher information metric. We will exploit these nice properties to define an isometric embedding from $\PP$ to ${\cal{E}}$ detailed in~(\ref{eq:IsometryEquality}). Then, we use this embedding  to construct a Gaussian process on PDFs  with a proper covariance function~(\ref{eq:cov:C:K}) and a predictive model~(\ref{predictor}).\\
 We first note that
the space of PDFs defined over  $\Omega$  with values in $\real_{+}$  can be viewed in different manners. The case where $\Omega$ is finite and the statistical model is parametric has been largely studied in the literature~\cite{Amari-book,Nihat-book}. In contrast, if $\Omega$ is infinite which is the case here, different modeling options have been explored~\cite{Friedrich91,Pistone-95,Anuj-book-2016,Bauer-2016}. We start with the ideas developed in~\cite{Friedrich91,shishido2005,Srivastava-2007,Nihat-book}  where  $\mathcal{P}$ is an infinite dimensional smooth manifold. That is, $\mathcal{P}$ is the space of probability measures that satisfy the normalization constraint.  Since we are interested in  statistical PDFs analysis on $\PP$, we need some geometrical tools~\cite{Helgason1978,Hyperbolic-Lee},\eg   geodesic. For the rest of the paper,  we view $\PP$ as a smooth manifold~(\ref{eq:P}) and we  impose a Riemannian structure on it with the Fisher-Rao metric~(\ref{eq:Fisher-Rao}). Let 
\begin{eqnarray}
 \label{eq:P}
\mathcal{P}=\{p:\Omega \to \mathbb{R} \hspace{0.05cm} | \hspace{0.05cm} p \geq 0 \hspace{0.1cm}\text{and} \int_{\Omega} p(t) dt=1 \}.
\end{eqnarray}
 be the space of all PDFs (positive almost everywhere) including nonparametric models.  We  identify any tangent space of $\mathcal{P}$, locally at each  $p$,  by 
 \begin{eqnarray}
 \label{eq:TangentP}
 T_p(\mathcal{P})=\{f:\Omega \to \mathbb{R} \hspace{0.05cm} | \hspace{0.05cm}  \int_{\Omega} f=0 \}
 \end{eqnarray}
As detailed in~\cite{Cencov-book-1982,Friedrich91,Bauer-2016}, the tangent space contains functions that are infinitesimally differentiable. But following~\cite{Helgason1978}, we have a constructive method of great importance that  allows one to  form a local version of any arbitrary $f$ that is continuously differentiable in a small neighborhood and null outside. Now that we have a smooth manifold and its tangent space, we can introduce a Riemannian metric.  This choice is very important since it will determine the structure of $\PP$ and consequently the covariance function of the Gaussian process. More details about the importance of the metric and the induced Riemannian structure are discussed in~\cite{Mitsuhiro-Geodesic-2015,Samir-FoCom12,bauer-metrics}. We also define and denote by $\PP_+$  the interior of $\mathcal{P}$. For the following, we consider without justification that any probability density  can be locally perturbed to be smooth enough~\cite{Helgason1978}. This is true in finite dimensional cases but the generalization to infinite dimensional cases is not straightforward.  Among several metrics, we are particularly interested in the Fisher-Rao metric defined, for any tangent vectors $f_1,f_2 \in T_p(\mathcal{P})$, by
\begin{eqnarray}
\label{eq:Fisher-Rao}
<f_1,f_2>_p=\int_{\Omega} \frac{f_1(t) f_2(t)}{p(t)} dt.
\end{eqnarray}
 Although this metric has nice properties  with an increasing interest~\cite{shishido2005,Amari-Rao-87,Cencov-book-1982},  $\PP$ equipped with $<.,.>_p$ is still numerically intractable. Therefore,  instead of working on $\PP$ directly, we consider a mapping  from $\PP$ to the Hilbert upper-hemisphere (positive part)  around the unity $1_{\PP}$ such that $1_{\PP}(t)=1$ for all $t$ in $\Omega$~\cite{Paper-Bhattacharyya43}. Thus, we  exploit the Riemannian isometry between $\PP$ and the  upper-hemisphere to extend  the notion of GPs to  the space of PDFs. Indeed,  we first define the map 
\begin{eqnarray}
\label{eq:SRDF}
\Psi  : \cal{P} & \rightarrow &   \cal{H} \\
p &\mapsto & \phi=2\sqrt{p}, \quad (\mathtt{p=\Psi^{-1}(\phi)=\frac{1}{4}\phi^2}) \nonumber
\end{eqnarray}
where
 \begin{eqnarray}
 \label{eq:HilbertSphareH} 
 \mathcal{H} = \{ \phi: \Omega \to \mathbb{R} \hspace{0.05cm} | \hspace{0.05cm}
  \phi  \geq 0 \hspace{0.1cm}
   \text{and}
   \int_{\Omega} \phi(t)^2 dt = 4  \}.
 \end{eqnarray}
 Note that $\phi$ is well defined since $p$ is nonnegative and 
 $\Psi$ is a  Riemannian isometry from $\cal{P}_+$ to $\cal{H}$ without the boundary~\cite{Hyperbolic-Lee}.   On the other hand, any element $\phi \in \HH$ can be uniquely projected as $\frac{1}{2} \phi$ to have a unit norm.  For simplicity and without loss of generality, we interpret $\cal{H}$ as the elements of unit Hilbert upper-hemisphere $\Sphere_+^{\infty}$ up to a multiplicative factor ($2$ here). From that point of view, we  identify $\cal{H}$ with $\Sphere_+^{\infty}$ and we define $\Psi(1_{\mathcal{P}})=1_{\mathcal{H}}$  to be the "unity pole" on $\cal{H}$. Note that $1_{\mathcal{H}}$ as the image of the uniform pdf $1_{\mathcal{P}}$ is a fixed point, \ie   $1_{\mathcal{H}}=\sqrt{1_{\mathcal{P}}}=1_{\mathcal{P}}$.  In this setup, we have 
\begin{eqnarray}
\|\phi \|_2^2 =\int_{\Omega} \phi(t)^2 dt=1,
\end{eqnarray}
 for any $\phi$ in $\cal{H}$  which allow us to consider $\cal{H}$,  when equipped with the integral inner product $<.,.>_{2}$, as the
unit  upper-hemisphere (positive part). Furthermore, for  arbitrary directions $f_1, f_2$ in $T_p(\cal{\PP})$ the Fisher-Rao metric as defined in~({eq:Fisher-Rao}) becomes  $<.,.>_{2}$ as follows:
\begin{eqnarray}
\label{eq:FR2L2}
<f_1,f_2>_p= < D_{f_1} \Psi ,D_{f_1} \Psi>_2.
\end{eqnarray}
with $D_{f_i} \Psi(p)(t) = \frac{f_i(t)}{\sqrt{p(t)}}$ for all $t \in \Omega$ and $i=1,2$.
 One of the main  advantages of this formulation is to exploit the nice properties of the unit Hilbert sphere such as geodesic, exponential map, log map, and the parallel transport. For the rest of the paper,  the geodesic distance $d_{\cal P}(p_1,p_2)$ between two PDFs $p_1$ and $p_2$ in $\cal P$, under the Fisher-Rao metric, is given by the geodesic distance $d_{\cal H}(\phi_1,\phi_2)$ (up to a factor 2) between their corresponding $\phi_1$ and $\phi_2$ on $\mathcal{H}$. We remind that the arc-length (geodesic distance) between distinct and non antipodal $\phi_1$ and $\phi_2$ on $\mathcal{H}$ is the angle   
$\beta=\arccos \left( <\phi_1,\phi_2>_2 \right)$. We also remind some  geometric tools that will be needed for  next sections as a lemma:
\begin{lem}
    With $\HH$ defined from~(\ref{eq:HilbertSphareH}) with unit norm  and  $T_{\phi}(\cal{H} )$ its tangent space at $\phi$,  we have the following:  
\begin{itemize}
	\item The exponential map is a bijective isometry from  the tangent space  $T_{\phi}(\cal{H})$ to 
	$\cal{H}$. For any 
 $w \in T_{\phi}(\cal{H})$, we write 
\begin{eqnarray}
    \label{eq:ExpMap}
\Exp{\phi}{w}=\cos(\|w\|_2)\phi + \sin(\|w\|_2)  \frac{w}{\|w\|_2}. 
\end{eqnarray}
	\item Its inverse, the log map 
 is defined from $\mathcal{H}$ to $T_{\phi_1}(\cal{H} )$ as 
	\begin{eqnarray}
    \label{eq:LogMap}
	\Log{\phi_1}{\phi_2}= \frac{\beta}{\sin(\beta)}(\phi_2-\cos(\beta)\phi_1). 
	\end{eqnarray} 
    \item 
    For any two elements $\phi_1$ and $\phi_2$ on $\HH$ the map $\Gamma: T_{\phi_1}(\HH) \rightarrow T_{\phi_2}(\HH)$
    parallel transports a vector $w$ from $\phi_1$ to $\phi_2$ and is given by:
	\begin{eqnarray}
\label{eq:ParallelTransport}
 \Gamma_{\phi_1 \rightarrowtail \phi_2}(w)= w-2\frac{(\phi_1+\phi_2)}{||\phi_1+\phi_2||_2^2}<w,\phi_2>_2
 \end{eqnarray}
\end{itemize}
\end{lem}
For more details, we refer  to~\cite{Hyperbolic-Lee}. As a special case,  we consider  the unity pole $\phi=1_{\cal H}$ and we denote ${\cal{E}}=T_{1}(\cal{H})$ the tangent space of $\cal{H}$ at $1_{\cal H}$. For simplicity, we note $\Log{1}{.}$ the log map from $\cal H$ to ${\cal{E}}$ and $\Exp{1}{.}$ its inverse. This choice is motivated by two reasons:  The numerical implementation and the fact that $1_{\HH}$ is the center of the geodesic disc $[0,\frac{\pi}{2}[$. Indeed and since all elements are on the positive part, the exponential map and its inverse are diffeomorphisms. So, one can consider any point on $\HH$ instead of $1_{\HH}$ to define the tangent space, \eg the Fr\'echet  mean. However, this is without loss for the numerical precision. Furthermore we can use the properties of the log map to show that: 
\begin{equation}
\label{eq:IsometryEquality}
||\Log{1}{\phi_i}-\Log{1}{\phi_j}||_2=d_{\cal H}(\phi_i,\phi_j)=\frac{1}{2}d_{\cal P}(p_i,p_j)
\end{equation}
  for any two $p_i$, $p_j$ on $\cal P$. Note that the multiplicative factor $\frac{1}{2}$ is important to guarantee the isometry but will not have any impact on the covariance function defined in~(\ref{eq:cov:C:K}) as it is implicit in the hyperparameter. 

\section{Gaussian Processes on PDFs}
\label{sec:GP}
In this section, we focus on constructing GPs on $ \PP$.
A GP $Z$ on $\PP$ is a random field indexed by $ \PP$ so that $(Z( p_1),\ldots,Z( p_n))$ is a multivariate Gaussian vector for any $n \in \mathbb{N}\backslash\lbrace{0}\rbrace$ and $p_1, \ldots , p_n \in   \PP$. 
A GP is completely specified by its mean function and its covariance  function. We define a mean function $m:   \PP \to \mathbb{R}$ and the covariance function 
$C:  \cal{P} \times \cal{P} \to \mathbb{R}$
 of a real process $Z$ as
\begin{align}
m(p_i)=&\mathbb{E} \big[Z(p_i)\big].  \\
C(p_i,p_j)=&\mathbb{E} \big[(Z(p_i)-m(p_i)) (Z(p_j)-m(p_j)) \big].
\end{align}
Thus, if a GP is assumed to have zero mean function ($m \equiv 0$), defining the covariance function completely defines the process behavior.
In this paper, we assume that the GPs are centered and we only focus on the issue of constructing a proper covariance function $C$  on  $ \cal{P}$. 
	\subsection{Constructing covariance functions on $ \cal{P}$}
	\label{sec:Covariance}
\noindent 	A covariance function $C$ on $ \cal{P}$ must satisfy the following conditions. For any $n \in \mathbb{N}\backslash\lbrace{0}\rbrace$ and $p_1, \ldots , p_n \in  \cal{P}$, the matrix $\mathbf{C}=[C(p_i,p_j)]_{i,j=1}^{n}$ is symmetric nonnegative definite. Furthermore, $C$ is called non-degenerate when the above matrix is invertible whenever $p_1,\ldots,p_n$ are two-by-two distinct~\cite{bachoc2017gaussian}. 
\textcolor{black}{The strategy that we adopt to construct covariance functions is to exploit the full isometry  map $\Log{1}{.}$ to  $\mathcal{E}$ given in~(\ref{eq:IsometryEquality}). That is, we construct covariance functions  of the form
	\begin{equation} 
	\label{eq:cov:C:K}
	C( p_i ,  p_j ) = 
	K( \| \Log{1}{ \phi_i } - \Log{1}{ \phi_j } \|_2 ),
	\end{equation}
	where  $K:\mathbb{R}^+ \to \mathbb{R}$.} 

	\begin{Prop} \label{prop:preservation:invertible}
		Let
		$K :\mathbb{R}^{+} \rightarrow \mathbb{R}$ be such that $K (u_i,u_j) = K( \|u_i - u_j\|_2 )$ is a covariance function on $\cal{E}$ and $C$ as defined as in~(\ref{eq:cov:C:K}). Then  
        \begin{enumerate}
            \item $C$ is a  covariance function.
		\item If  $[K(\|u_i-u_j\|_2)]_{i,j=1}^{n}$ is invertible, then $C$ is 
		non-degenerate.
    \end{enumerate}
	\end{Prop}
A closely related proof when dealing with Cumulative Density Functions (CDFs)
is given in~\cite{bachoc2017gaussian}. 
	In practice, we can select the function $K$ from the Mat\'ern family, letting for $t \geq 0$
 \begin{eqnarray}
 K_{\theta}(t)
 =
 \frac{\delta^2 }{ \Gamma(\nu)2^{\nu-1}} \Big(\frac{2 \sqrt{\nu t}}{\alpha} \Big)^\nu K_\nu \Big(\frac{2 \sqrt{\nu t}}{\alpha} \Big), 
 \end{eqnarray}
\textcolor{black}{where $K_\nu$ is a modified Bessel function of the second kind and $\Gamma$ is the gamma function.
We note
 $\theta=(\delta^2,\alpha,\nu) \in
	\Theta$
	where $\delta^2>0$ is the variance parameter,
	$\alpha>0$ is the correlation length parameter
	and $\nu=\frac{1}{2}+k (k \in \mathbb{N})$ is the smoothness parameter.
	The Mat\'{e}rn form~\cite{Ste1999} has the desirable property that GPs have realizations (sample paths) that are $k$ times differentiable~\cite{genton2015}, which prove its smoothness as function of $\nu$. 
	As $\nu \to \infty$, the Mat\'{e}rn covariance function approaches the squared exponential  form, whose realizations are infinitely differentiable. For $\nu = \frac{1}{2}$, the Mat\'{e}rn takes the exponential form.
	From Proposition~\ref{prop:preservation:invertible}, the Mat\'ern covariance function defined by $C(p_i , p_j) = K_{\theta}( \| \Log{1}{ \phi_i } - \Log{1}{ \phi_j } \|_2 )$ is indeed  non-degenerate.}
\subsection{Regression on $\mathcal{P}$}
\label{sec:Regression}
\noindent
Having set out the conditions on the covariance function,  we can define the  regression model on $\cal{P}$ by 
\begin{eqnarray}
\label{reg-model}
y_i=Z(p_i) + \epsilon_i, \hspace{0.2cm} i=1,\dots,n,
\end{eqnarray} where $Z$ is a zero mean GP indexed by $\mathcal{P}$
with a covariance function in the set $\{C_{\theta}; \theta \in \Theta\}$ and $\epsilon_i 
\overset{\text{iid}}{\sim}
\mathcal{N}(0,\gamma^2)$.
Here $\gamma^2$ is the observation noise variance, that we suppose to be known for simplicity.
Moreover, 
we note $\mathbf{y}=(y_1,\dots,y_n)^T$,
$\mathbf{p}=(p_1,\dots,p_n)^T$ and $\mathbf{v}=(v_1,\dots,v_n)^T=
(\Log{1}{\Psi(p_1)},\dots,\Log{1}{\Psi(p_n)})^T$.
The likelihood term is $\mathbb{P}(\mathbf{y}|Z(\mathbf{p}))=\mathcal{N}(Z(\mathbf{p}),\gamma^2 I_n)$ where 
$I_n$ is the identity matrix. Moreover,
 the prior on $Z(\mathbf{p})$ is $\mathbb{P}(Z(\mathbf{p}))=\mathcal{N}(0,\mathbf{C}_{\theta})$ with
$\mathbf{C}_{\theta}=[K_{\theta}(\|v_i-v_j\|_2)]_{i,j=1}^{n}$. 
We use the product of likelihood and  prior terms to perform the integration yielding the log-marginal likelihood
\begin{eqnarray}
l_{r}(\theta) =
-\mathbf{y}^T (\mathbf{C}_{\theta}+\gamma^2 I_n)^{-1} \mathbf{y}
-\log |\mathbf{C}_{\theta}+\gamma^2 I_n|  -\frac{n}{2} \log 2\pi. 
\label{likelihood-reg}
\end{eqnarray}
Let $\theta=\{\theta^j\}_{j=1}^{3}=(\delta^2,\alpha,\nu)$  denote the parameters of the Mat\'ern covariance function  $K_{\theta}$.
The partial derivatives of $l_{r}(\theta)$ with respect to $\theta^j$ are
\begin{eqnarray}
\label{grad1}
\frac{\partial l_{r}(\theta)}{\partial \theta^j}=\frac{1}{2} \mathbf{y}^T \mathbf{C}_{\theta}^{-1} \frac{\partial \mathbf{C}_{\theta}}{\partial \theta^j} \mathbf{C}_{\theta}^{-1} \mathbf{y} - \text{tr}\big[\mathbf{C}_{\theta}^{-1}\frac{\partial \mathbf{C}_{\theta}}{\partial \theta^j}\big].
\end{eqnarray}
For an unobserved PDF $p^*$ and by deriving the conditional distribution, we arrive at the key predictive equation 
\begin{eqnarray}
\mathbb{P}(Z(p^*)|\mathbf{p},\mathbf{y},p^*)= \mathcal{N}(\mu(p^*),\sigma^2(p^*)),
\end{eqnarray}
with
\begin{equation}
\left\lbrace
\begin{aligned}
\mu(p^*)&={\mathbf{C}_{\theta}^*}^T (\mathbf{C}_{\theta}+\gamma^2 I_n)^{-1} \mathbf{y},\\
\sigma^2(p^*)&=C_{\theta}^{**} - 
{\mathbf{C}_{\theta}^*}^T(\mathbf{C}_{\theta}+\gamma^2 I_n)^{-1} \mathbf{C}_{\theta}^*,\\
\end{aligned}
\right.
\end{equation}
where $\mathbf{C}^*_{\theta}=K_{\theta}(\mathbf{v},v^*)$ and 
$C_{\theta}^{**}=K_{\theta}(v^*,v^*)$ for $v^*=\Log{1}{\Psi(p^*)}$.
As we have introduced GP regression indexed by PDFs, we will present GP classifier in the next section. 
\subsection{Classification on $\mathcal{P}$}
\label{sec:Classification}
\noindent For the classification part, we focus on the case  of  binary outputs, i.e., $y_i \in \{-1,+1\}$.
We first adapt the Laplace approximation to GPc indexed by PDFs in Section~\ref{sec:laplace}. We also give the approximate marginal likelihood and the Gaussian predictive distribution in Section~\ref{sec:marginal}.
\subsubsection{Approximation of the posterior}
\label{sec:laplace}
The likelihood is the product of individual likelihoods 
$
\mathbb{P}(\mathbf{y}|Z(\mathbf{p}))=\prod_{i=1}^{n} \mathbb{P}(y_i|Z(p_i))
$
where $\mathbb{P}(y_i|Z(p_i))=\sigma(y_i Z(p_i))$ and $\sigma(.)$ refers to the  sigmoid function satisfying $\sigma(t)=\frac{1}{1+\exp(-t)}$. As for regression, the prior law of GPc is 
$
\mathbb{P}(Z(\mathbf{p}))=\mathcal{N}(0,\mathbf{C}_\theta)
$.
From the Bayes' rule, the posterior distribution of $Z(\mathbf{p})$ satisfies
\begin{eqnarray}
\label{posterior}
\mathbb{P}(Z(\mathbf{p})|\mathbf{y})  = 
\frac{\mathbb{P}(\mathbf{y}|Z(\mathbf{p})) \times \mathbb{P}(Z(\mathbf{p}))}{\mathbb{P}(\mathbf{y}|\mathbf{p},\theta)}, 
\propto \mathbb{P}(\mathbf{y}|Z(\mathbf{p})) \times \mathbb{P}(Z(\mathbf{p})), 
\end{eqnarray}
where $\mathbb{P}(\mathbf{y}|\mathbf{p},\theta)$ is the exact marginal likelihood.
The log-posterior is simply proportional to $
 \log \mathbb{P} (\mathbf{y}|Z(\mathbf{p})) -  \frac{1}{2} Z(\mathbf{p})^T \mathbf{C}_{\theta}^{-1} Z(\mathbf{p})$.
For the Laplace approximation, we approximate the posterior  given in~(\ref{posterior})
 by a Gaussian distribution. We can find the maximum a posterior (MAP) estimator denoted by $\hat{Z}(\mathbf{p})$, iteratively, according to
\begin{eqnarray}
Z^{k+1}(\mathbf{p})=(\mathbf{C}_{\theta}+\mathbf{W})^{-1} (\mathbf{W} Z^{k}(\mathbf{p}) + \nabla \mathbb{P} (\mathbf{y}|Z^k(\mathbf{p})) ),
\end{eqnarray}
where $\mathbf{W}$ is  a $n \times n$ diagonal matrix with entries
$\mathbf{W}_{ii}=\frac{\exp(-\hat{Z}(p_i))}{(1+\exp(-\hat{Z}(p_i)))^2}$.
Using the MAP estimator, we can specify the Laplace approximation of the posterior by 
\begin{eqnarray}  
\hat{\mathbb{P}}(Z(\mathbf{p})|\mathbf{p},\mathbf{y})= \mathcal{N}(\hat{Z}(\mathbf{p}),(\mathbf{C}_{\theta}^{-1}+\mathbf{W})^{-1}).
\end{eqnarray}
\subsubsection{Predictive distribution}
\label{sec:marginal}
We evaluate the approximate marginal likelihood denoted by $\hat{\mathbb{P}}(\mathbf{y}|\mathbf{p},\theta)$ 
instead of the exact marginal likelihood  $\mathbb{P}(\mathbf{y}|\mathbf{p},\theta)$ given in the denominator of~(\ref{posterior}). 
Integrating out  $Z(\mathbf{p})$, the log-marginal likelihood is approximated by
\begin{eqnarray}
\label{log-marginal}
l_{c}( \theta)=-\frac{1}{2} \hat{Z}(\mathbf{p})^T {\mathbf{C}}_{\theta}^{-1} \hat{Z}(\mathbf{p})  + \log p (\mathbf{y}|\hat{Z}(\mathbf{p})) 
 - \frac{1}{2} \log \big|I_n +\mathbf{W}^{\frac{1}{2}} \mathbf{C}_{\theta} \mathbf{W}^{\frac{1}{2}}\big|. 
\end{eqnarray}
The partial derivatives of $l_c(\theta)$
with respect to $\theta^j$
satisfy
\begin{eqnarray}
\label{grad2}
\frac{\partial l_{c}(\theta)}{\partial \theta^j}=\frac{\partial l_{c}(\theta)}{\partial \theta^j}|_{\hat{Z}(\mathbf{p})}+\sum_{i=1}^{n} \frac{\partial l_{c}(\theta)}{\partial \hat{Z}(p_i)} \frac{\partial \hat{Z}(p_i)}{\partial \theta^j}.
\end{eqnarray}
The first term, obtained when we assume that $\hat{Z}(\mathbf{p})$ (as well as $\mathbf{W}$) does not depend on $\theta$, satisfies
\begin{eqnarray}
\frac{\partial l_{c}(\theta)}{\partial \theta^j}|_{\hat{Z}(\mathbf{p})} = 
\frac{1}{2} {\hat{Z}(\mathbf{p})}^{T} \mathbf{C}_{\theta}^{-1} \frac{\partial \mathbf{C}_{\theta}}{\partial \theta^j} \mathbf{C}_{\theta}^{-1} \hat{Z}(\mathbf{p}) 
 -\frac{1}{2} \text{tr}\big[(\mathbf{C}_{\theta}+{\mathbf{W}}^{-1})^{-1} \frac{\partial \mathbf{C}_{\theta}}{\partial \theta^j} \big]. 
\end{eqnarray}
The second term, obtained when we suppose that only $\hat{Z}(\mathbf{p})$ (as well as $\mathbf{W}$) depends on $\theta$, is determined by 
\begin{eqnarray}
\frac{\partial l_{c}(\theta)}{\partial \hat{Z}(p_i)}=-\frac{1}{2}\big[(\mathbf{C}_\theta^{-1} + \mathbf{W})^{-1}   \big]_{ii} 
\frac{\partial^3 \log p(\mathbf{y}|\hat{Z}(\mathbf{p}))}{\partial^3  \hat{Z}(p_i) },
\end{eqnarray}
and 
\begin{eqnarray}
\frac{\partial \hat{Z}(\mathbf{p})}{\partial \theta^j}=\big( I_n + \mathbf{C}_{\theta} \mathbf{W} \big)^{-1} 
\frac{\partial \mathbf{C}_{\theta}}{\partial \theta^j} \nabla \log p (\mathbf{y}|\hat{Z}(\mathbf{p})).
\end{eqnarray}
Given an unobserved PDF $p^*$, the predictive distribution at $Z(p^*)$ is given by 
\begin{eqnarray}
\hat{\mathbb{P}}(Z(p^*)|\mathbf{p},\mathbf{y},p^*)= \mathcal{N}(\mu(p^*),\sigma^2(p^*)),
\end{eqnarray}
with 
\begin{equation}
\left\lbrace
\begin{aligned}
\mu(p^*)&={\mathbf{C}_{\theta}^*}^T \mathbf{C}_{\theta}^{-1} \hat{Z}(\mathbf{p}),\\
\sigma^2(p^*)&=C_{\theta}^{**} - 
{\mathbf{C}_{\theta}^*}^T(\mathbf{C}_{\theta}+\mathbf{W}^{-1})^{-1} \mathbf{C}_{\theta}^*.\\
\end{aligned}
\right.
\end{equation}
Finally, using the moments of prediction,  the predictor
for $y^*=+1$ is 
\begin{eqnarray}
\pi(p^*)= \int_{\mathbb{R}}
\sigma(Z^*) \hat{\mathbb{P}}(Z^*|\mathbf{p},\mathbf{y},p^*) d Z^*,
\label{predictor}
\end{eqnarray}
where we note $Z^*=Z(p^*)$ for simplicity.
\subsection{Covariance parameters estimation}
\noindent  \textcolor{black}{The  marginal likelihoods for both regression and classification  depend on the covariance parameters controlling the stationarity of the GP. To show potential applications  of this framework, 
	we  explore several optimization methods 
	in Section~\ref{section:gradient}
	 and Section~\ref{section:hmc}.} 
\subsubsection{log-marginal likelihood gradient}
\label{section:gradient}
\textcolor{black}{In the marginal likelihood estimation,  the parameters
are obtained by maximizing the log-marginal likelihood with respect to $\theta$}, i.e., finding
\begin{eqnarray}
\hat{\theta}=\argmax_{\theta} l_{l}(\theta),
\end{eqnarray}
\textcolor{black}{where $l_{l}(\theta)$ is given  in~(\ref{likelihood-reg}) by  $l_{r}(\theta)$  
 for regression or 
 $l_{c}(\theta)$  in(~\ref{log-marginal}) 
 for classification. We summarize the main steps in Algorithm~\ref{algo1}}.	
 \\[1 cm]
\textcolor{black}{	
\begin{algorithm}[H]
		\caption{Gradient descent.}
		\begin{algorithmic}[1]
			\REQUIRE log-marginal likelihood $l_l$ and its gradient  $\nabla l_l$ 
			\ENSURE $\hat{\theta}$ \\
			\REPEAT 
			\STATE $\nabla l_{l}(\theta(k))=\{ \frac{\partial l_{l}(\theta(k))}{\partial \theta^j}\}_{j=1}^3$
			from~(\ref{grad1})
		 or~(\ref{grad2})
			\STATE Find the step-size $\lambda$
			(e.g., by backtracking line search)
			\STATE Evaluate $\theta(k+1)=\theta(k)-\lambda \nabla l_{l}(\theta(k))$
			\STATE Set $k=k+1$
			\UNTIL{$||\nabla l_{l}||_2$  is small enough or a maximum  iterations is reached} 
		\end{algorithmic}
		\label{algo1}
	\end{algorithm}
}

\subsubsection{HMC sampling}
\label{section:hmc}
\textcolor{black}{Generally, the marginal likelihoods are non-convex functions. Indeed, conventional optimization routines may not find the most probable candidate  leading to a lost of
robustness and uncertainty quantification.
To deal with such limitations, we use weak prior distributions  for $\delta^2$ and $\alpha$ whereas $\nu$ is simply estimated by cross-validation~\cite{Neal97montecarlo}:  
\begin{eqnarray}
\mathbb{P}(\delta^2,\alpha)= \mathbb{P}(\delta^2) \times \mathbb{P}(\alpha),
\end{eqnarray}
with  $\delta^2$ and $\alpha$ being independent}. Following~\cite{Gelman06priordistributions},
$\delta^2$  will be assigned  a half-Cauchy (positive-only) prior, \ie  $\mathbb{P}(\delta^2)=\mathcal{C}(0,b_{\delta^2})$ and $\alpha$  an inverse gamma, \ie   $\mathbb{P}(\alpha)=\mathcal{IG}(a_\alpha,b_\alpha)$. 
Consequently,
the log-marginal posterior is proportional to 
\begin{eqnarray}
l_{p}(\delta^2,\alpha)= l_l(\theta) + \log \mathbb{P}(\delta^2) + \log \mathbb{P}(\alpha). 
\end{eqnarray}
When sampling from continuous variables, HMC can prove to be a more powerful tool than usual MCMC sampling. We define the Hamiltonian as the sum of a potential energy 
and a kinetic energy: 
\begin{eqnarray}
E((\theta^1,\theta^2),(s^1,s^2))&=&E^1(\theta^1,\theta^2)+E^2(s^1,s^2)
 -l_{p}(\theta^1,\theta^2)  + \frac{1}{2} \sum_{j=1}^{2} {s^j}^2, 
 \label{Hamiltonian}
\end{eqnarray}
which means that $(s^1,s^2) \sim \mathcal{N}(0,I_2)$.
Instead of sampling from $\exp\big(l_{p}(\theta^1,\theta^2)\big)$ directly, HMC operates by sampling from the distribution $\exp \big(- E((\theta^1,\theta^2), (s^1,s^2))\big)$. The differential equations are given by
\vspace{-.1cm}
\begin{equation}
\frac{d \theta^j}{d t}=\frac{\partial E}{\partial s^j}=s^j  \quad \text{and}  \quad 
\frac{d s^j}{d t}=-\frac{\partial E}{\partial \theta^j} = -\frac{\partial E^1}{\partial \theta^j}, 
\end{equation}
for $j=1,2$.
In practice, we can not simulate Hamiltonian dynamics exactly because of  time discretization.  To maintain invariance of the Markov chain, however, care must be taken to preserve the properties of volume conservation and time reversibility. The leap-frog algorithm, summarized in Algorithm~\ref{algo3}, maintains these properties~\cite{1206.1901}. 

	\begin{algorithm}[H]
		\caption{Leap-frog.}
		\begin{algorithmic}[1]
			\FOR{$k=1,2,\dots$}
			\STATE $ s^j (k+\frac{\lambda}{2}) = s^j (k) -\frac{\lambda}{2}
			\frac{\partial}{\partial \theta^j}  E^1(\theta^1(k),\theta^2(k))$
			where $\lambda$ is a finite step-size
			\STATE $\theta^j(k+\lambda)=\theta^j(k) + \lambda s^j (k+\frac{\lambda}{2}) $ 
			\STATE $ s^j (k+\lambda)=s^j (k+\frac{\lambda}{2}) -\frac{\lambda}{2}
			\frac{\partial}{\partial \theta^j} E^1(\theta^1(k+\lambda),\theta^2(k+\lambda)) $
			\ENDFOR
		\end{algorithmic}
		\label{algo3}
	\end{algorithm}

We thus perform a half-step update of the velocity at time $k+\frac{\lambda}{2}$, which is then used to compute $\theta^j(k + \lambda)$ and $s^j(k + \lambda)$. A new state $((\theta^1(N),\theta^2(N)),(s^1(N),s^2(N)))$ is then accepted with the probability 
\begin{eqnarray}
\label{acceptance}
\min \Big(1,\frac{\exp\big(-E((\theta^1(N),\theta^2(N),(s^1(N),s^2(N)) \big)}{\exp\big(-E((\theta^1(1),\theta^2(1),(s^1(1),s^2(1)) \big)}  \Big). 
\end{eqnarray}
We summarize the HMC sampling in Algorithm~\ref{algo4}.

	\begin{algorithm}[H]
		\caption{HMC sampling.}
		\begin{algorithmic}[1]
			\REQUIRE log-marginal posteriors $l_p$ and its gradient  $\nabla l_p$ 
			\ENSURE $\hat{\theta}$  
			\STATE Sample a new velocity from a  Gaussian distribution $(s^1(1),s^2(1)) \sim \mathcal{N}(0,I_2)$
			\STATE Perform $N$ leapfrog steps to obtain the new state $(\theta^1(N),\theta^2(N))$ and velocity $(s^1(N),s^2(N))$ from Algorithm~\ref{algo3} 
			\STATE  Perform accept/reject of $(\theta^1(N),\theta^2(N))$ with acceptance probability defined in~(\ref{acceptance}).
		\end{algorithmic}
		\label{algo4}
	\end{algorithm}

\section{Experimental Results}
\label{sec:results}
\noindent In this section, we test and illustrate the proposed methods using synthetic, semi-synthetic and real data. For all experiments, we study
the empirical results of a Gaussian process indexed by PDFs for both regression and classification.

\noindent \textbf{Baselines.}
We compare results of GP indexed by PDFs (GPP) where the parameters are estimated by gradient descend (G-GPP) and HMC (HMC-GPP) to:  Functional Linear Model (\textbf{FLM})~\cite{Ramsay-1991} for regression, Nonparametric Kernel Wasserstein (\textbf{NKW})~\cite{5513626} for regression,  
  A GPP based on the Wasserstein  distance (\textbf{W-GPP})~\cite{NIPS2017_7149,bachoc2017gaussian} for classification, and 
  a GPP based on the Jensen-Shannon divergence (\textbf{JS-GPP})~\cite{Nguyen-Div-JS} for classification.

\noindent \textbf{Performance metrics.}
For regression, we illustrate the performance of the proposed framework in terms of
root mean square error (RMSE) and negative  log-marginal likelihood (NLML). For classification, we consider 
 accuracy, area under curve (AUC) and NLML.

 \subsection{Regression}
\begin{figure}[t!]
    \centering
    \includegraphics[height=5cm,width=0.75\textwidth]{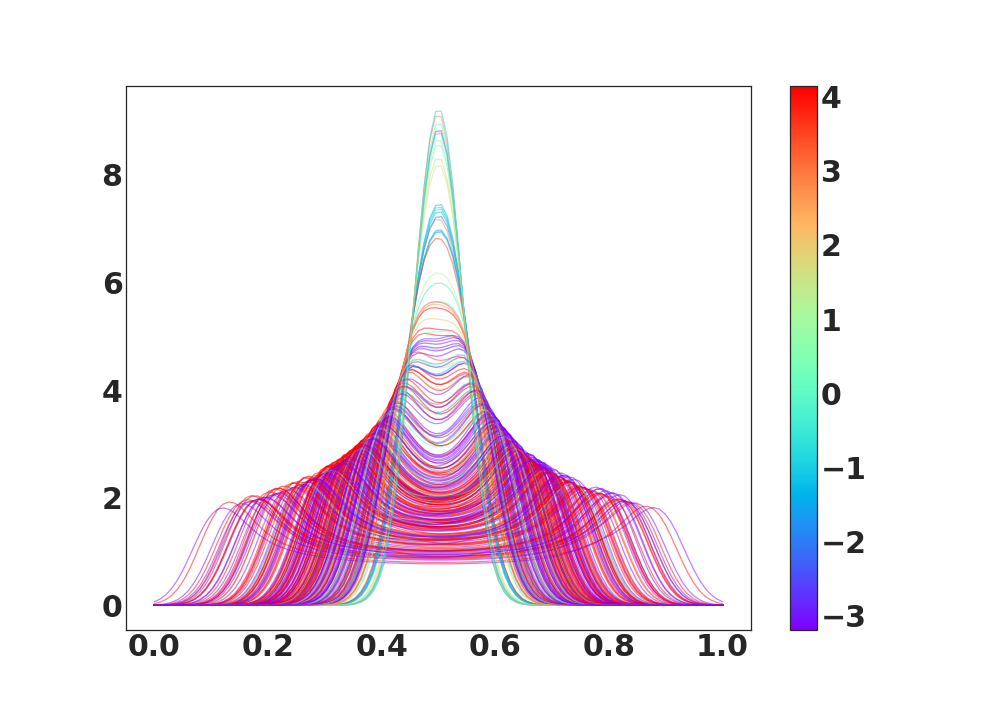}
    \caption{Examples of PDFs input for regression. The output with continuous value in $[-3,4]$ is illustrated by a colorbar.}
    \label{fig:DatasetRegression}
\end{figure}

\noindent	\textbf{Dataset.}	We first consider a synthetic dataset  where
we  observe a finite set of functions simulated according to ~(\ref{reg-model}) as 
  $Z(p_i)= h(<\sqrt{p_i},\sqrt{\tilde{p}}>_2)=0.5<\sqrt{p_i},\sqrt{\tilde{p}}>_2+0.5$.
 In this example, we consider
 a truncated Fourier basis (TFB) with random Gaussian coefficients
 to form the original functions satisfying
 $g_i(t)=\delta_{i,1}\sqrt{2} \sin(2 \pi t) + \delta_{i,2}\sqrt{2} \cos(2 \pi t)$ with $\delta_{i,1},\delta_{i,2} \sim \mathcal{N}(0,1)$.
 We also take  $\tilde{g}(t)=-0.5\sqrt{2} \sin(2 \pi t) + 0.5 \sqrt{2} \cos(2 \pi t)$.
We suppose that $\tilde{p}$ and $p_i$s refer to the corresponding PDFs of $\tilde{g}$ and $g_i$s
 estimated  from samples using the nonparametric kernel method (bandwidths were selected using the method given in~\cite{Botev-2010}). Examples of $n=100$ estimates are displayed in Fig.~\ref{fig:DatasetRegression} with  colors depending on their output levels.

\noindent	\textbf{Regression results.}  
Focusing on RMSE, we summarize all results   in Table~\ref{tab:RegressionRMSE}.
Accordingly,   the
proposed G-GPP gives better precision  than FLM. 
On the other hand, 
HMC-GPP substantially outperforms NKW with a significant margin.
As illustrated in Table~\ref{tab:RegressionAccuracy}, we note that the proposed  methods are more efficient than the baseline FLM when maximizing the log-marginal likelihood. Again, this is a very simple explanation  on how  the quality of GPP strongly depends on parameters estimation method.
In addition, G-GPP stated in Algorithm~\ref{algo1}  
is very effective from a computational point of view. 
\begin{table}[h!]
    \centering
    \caption{ Regression:  RMSE as a performance metric.}
    {\begin{tabular}{|c|c|c|c|c|c|c|c|}
            \hline
            \multicolumn{2}{|c|}{G-GPP} & \multicolumn{2}{|c|}{HMC-GPP} & \multicolumn{2}{|c|}{FLM} & 
            \multicolumn{2}{|c|}{NKW} \\
            \hline
            mean & std & mean & std & mean & std & mean & std
            \\
            \hline
            $\mathbf{0.07}$  &  0.03 & 0.13 & 0.31 & 0.10 &  0.04  & 0.28 & 0.01 \\
            \hline
    \end{tabular}}
    \label{tab:RegressionRMSE}	
\end{table}

\begin{table}[h!]
    \centering
    \caption{Regression: negative log-marginal likelihood as a performance metric.}
    {\begin{tabular}{|c|c|c|c|c|c|}
            \hline
            \multicolumn{2}{|c|}{G-GPP} & \multicolumn{2}{|c|}{HMC-GPP} & \multicolumn{2}{|c|}{FLM} \\
            \hline
            mean & std & mean & std & mean & std 
            \\
            \hline
            73.28	  &  1.14 & $\mathbf{21.89}$ & 5.32 & 329.66  &  6.52   \\
            \hline
    \end{tabular}}
    \label{tab:RegressionAccuracy}
\end{table}

\subsection{Classification}
In this section, we  perform some extensive experiments to evaluate the proposed methods  using a second category of datasets.

\subsubsection{Datasets for classification}
\begin{figure}[t!]
     \centering
     \setlength{\tabcolsep}{-20pt}
     \begin{tabular}{c@{}c@{}}
         \includegraphics[width=0.35\textwidth]{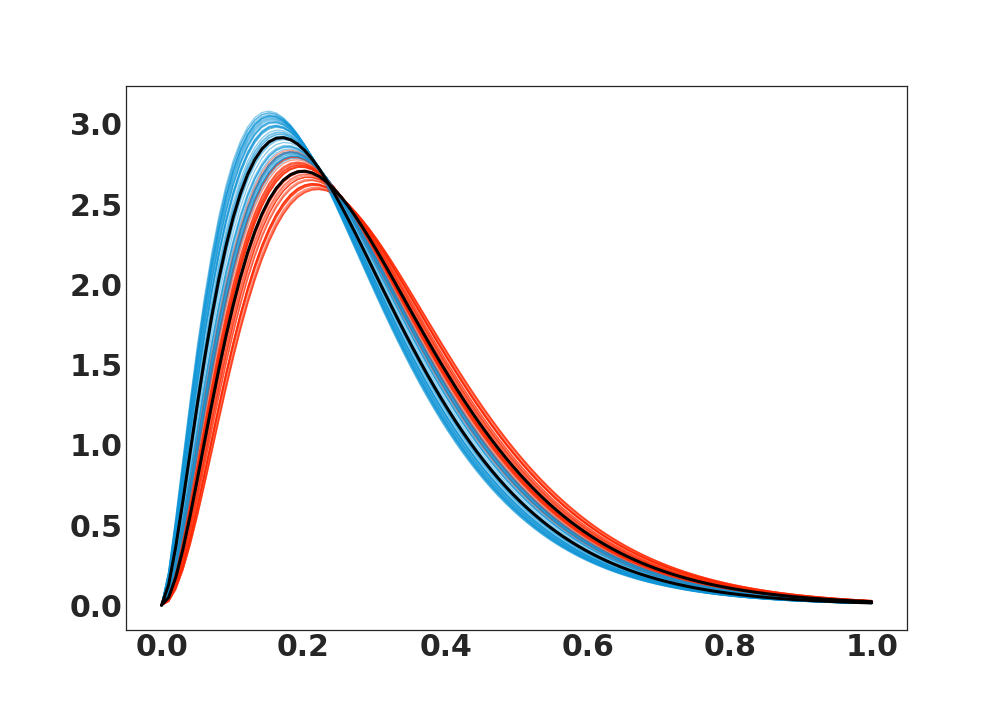}&
         \includegraphics[width=0.35\textwidth]{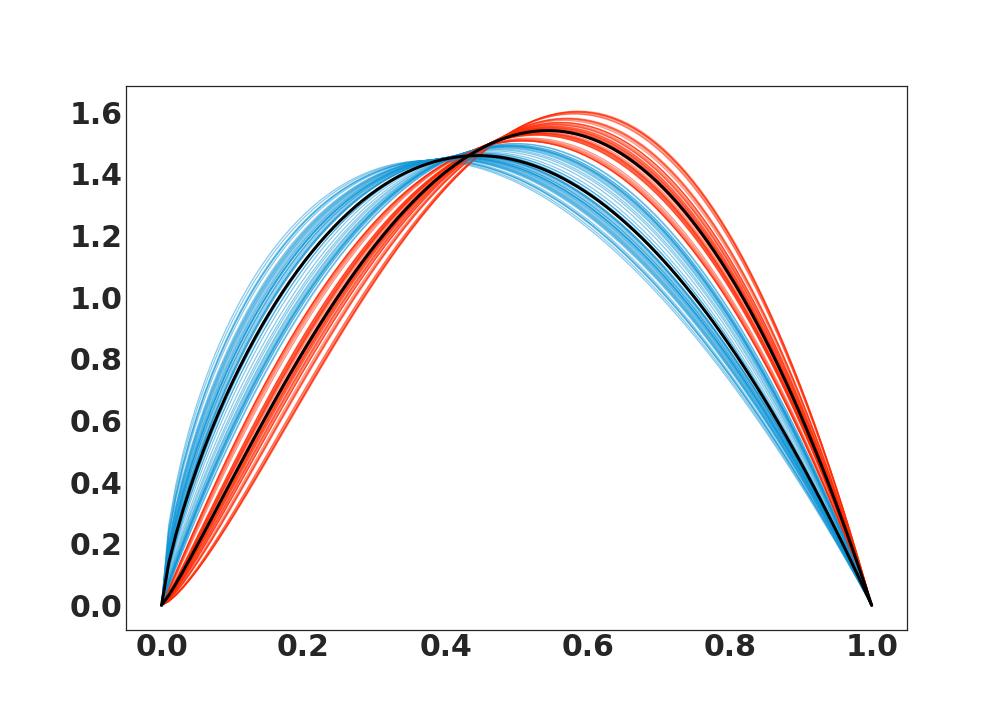}\\
         (a)  &	(b)  \\
     \end{tabular}
     \begin{tabular}{c@{}c@{}c@{}}
         \includegraphics[width=0.35\textwidth]{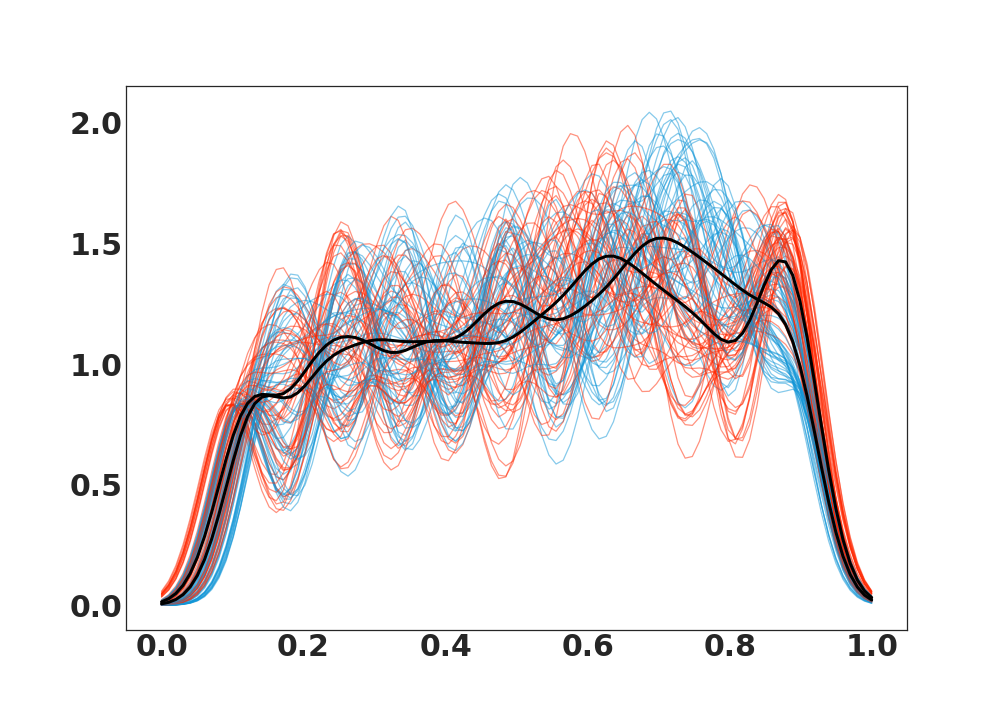} &
         \includegraphics[width=0.35\textwidth]{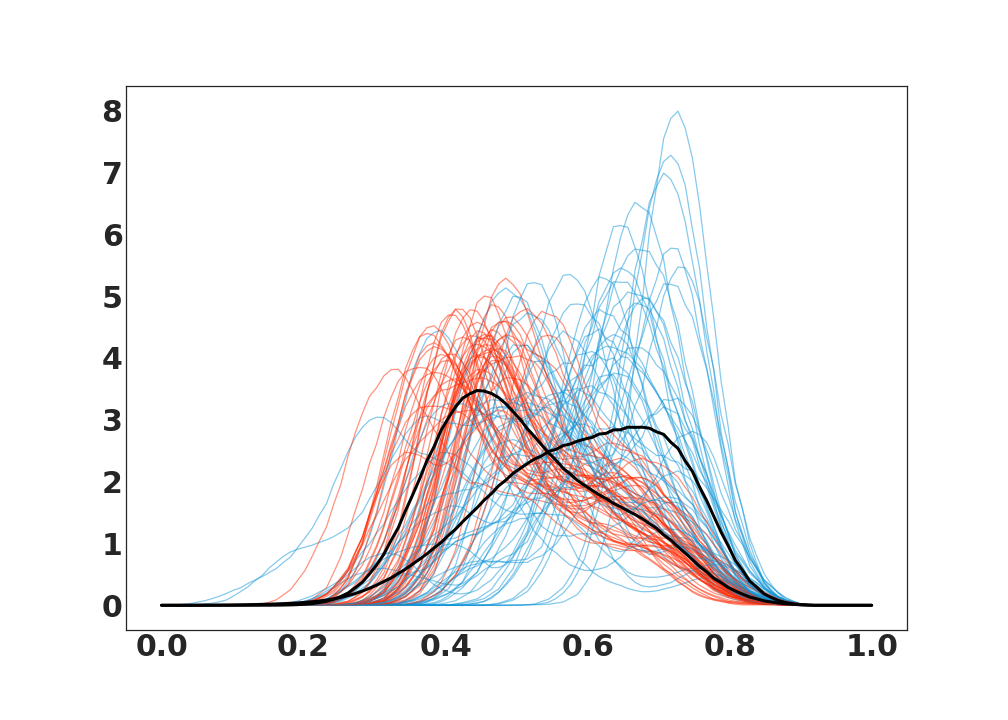}&
         \includegraphics[width=0.35\textwidth]{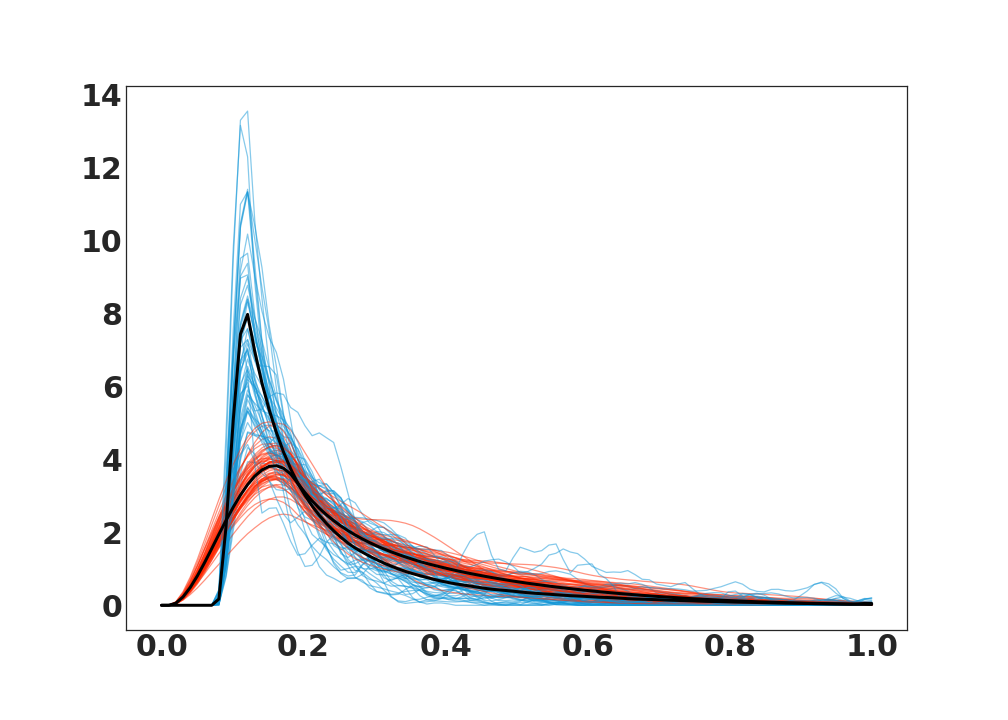}\\
       (c)  &	(d)  & (e)  \\
     \end{tabular}
     \caption{
         Synthetic PDFs for (a) InvGamma  and (b) Beta  with class 1 (red) and class 2 (blue).	Semi-synthetic PDFs for (c) Growth with girls (red) and boys (blue). 
         Real PDFs for (d) Temp with  uninfected (red) and  infected (blue).
         Real PDFs for (e) Plants with  disease (red) and healthy (blue).
         The Fr\'echet mean for each class in black.}
     \label{fig:DatasetsClassification}
 \end{figure}

\noindent \textbf{Synthetic datasets}.
We consider a dataset of two synthetic PDFs of beta and inverse gamma distributions. This choice is very crucial for many reasons   since beta is defined on  $[0, 1]$,  parametrized by two positive  parameters,  and has been widely used to represent a large family of PDFs with finite support in various fields. Increasingly, the inverse gamma plays an important role to characterize  random fluctuations affecting wireless channels~\cite{Wireless-Paper}. In both examples,  the covariance matrix with $\mathbb{L}^2$ distance and Total Variation TV-distance have a very low rank. 
We performed this experiment
by simulating $n=200$ pairs of PDFs slightly different for the two classes.  Each observation represents  a density  when we add a random white noise.  We refer to these datasets as Beta and InvGamma, see random examples in Fig.~\ref{fig:DatasetsClassification} (a\&b).
    We also illustrate the Fr\'echet mean  for each class. 
    The search of the mean is performed using a gradient approach detailed  in~\cite{Srivastava-2007}. \\
\noindent \textbf{Semi-synthetic dataset}. Data represent clinical growth charts for children from $2$ to $12$ years~\cite{Ramsay-1991}. We refer to this dataset as Growth. We simulate the charts from centers for disease control and prevention~\cite{Kuczmarski} through the available quantile values. The main   goal is   to classify observations by gender.
	Each simulation	represents the size growth (the increase)  of a child according to his age ($120$ months). We represent observations as 
 nonparametric PDF and we display some examples  in Fig.~\ref{fig:DatasetsClassification} (c). For each class: girls (red) and boys (blue) we show  the Fr\'echet mean  in  black.
 
 \noindent \textbf{ Real dataset}. The first   public dataset consists of  $1500$ images representing  maize leaves~\cite{DeChant} with   
 specific textures whereas the goal is to distinguish  healthy and non-healthy plants. We refer to this dataset as Plants.
 Motivated by this application, we first represent each image with its wavelet-deconvolved version and form a high-dimensional vector of  $262144$ components.
   \begin{figure}[h!]
     \centering
     \begin{tabular}{c@{}c@{}c@{}}
         \includegraphics[height=2.5cm,width=0.3\textwidth]{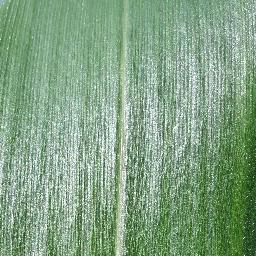} &
         \includegraphics[height=2.5cm,width=0.3\textwidth]{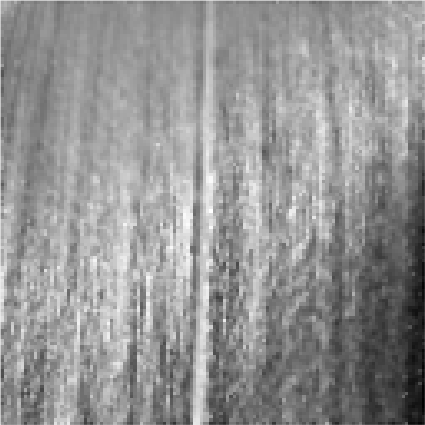}&
         \includegraphics[height=2.5cm,width=0.3\textwidth]{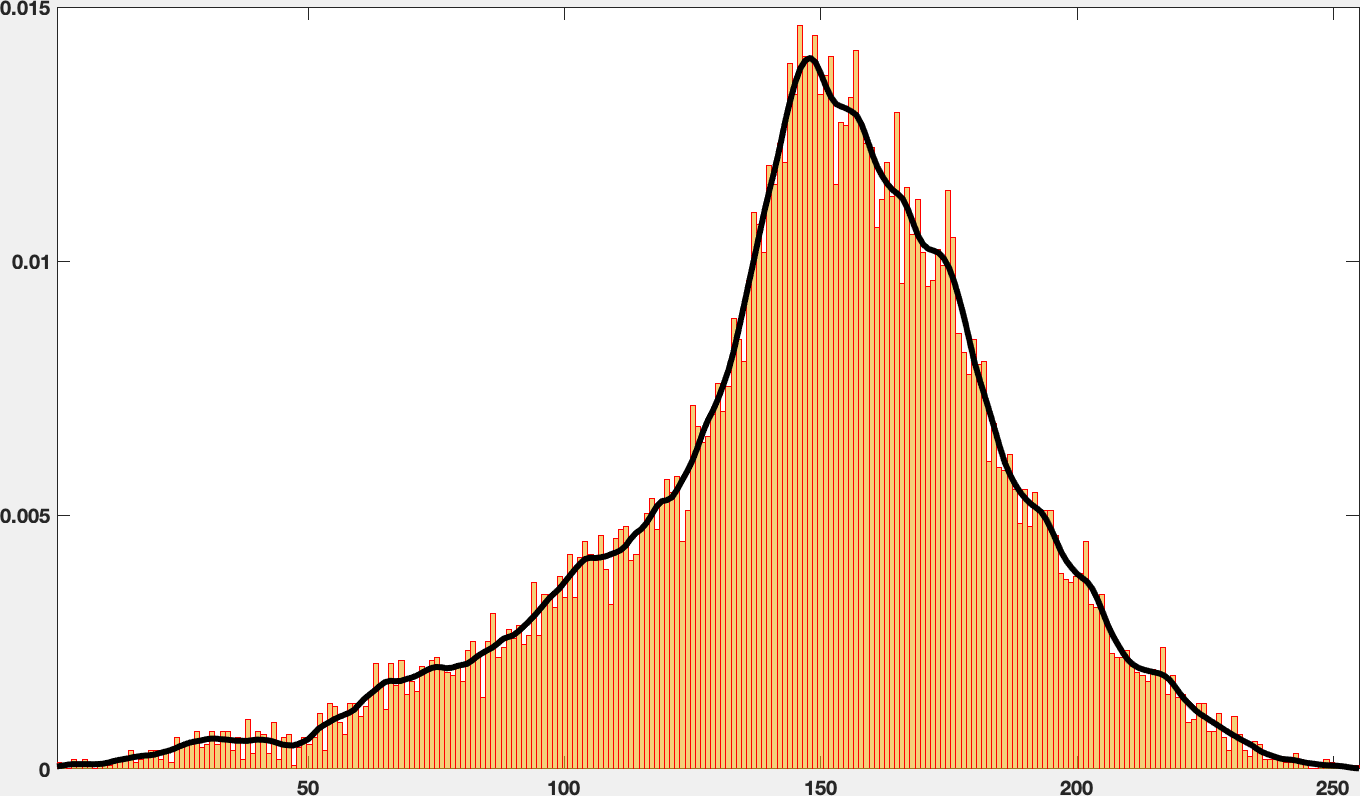}\\
         
         \includegraphics[height=2.5cm,width=0.3\textwidth]{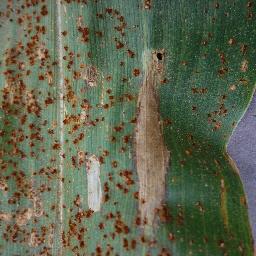} &
         \includegraphics[height=2.5cm,width=0.3\textwidth]{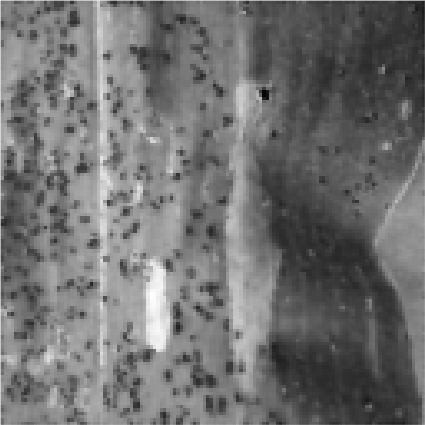}&
         \includegraphics[height=2.5cm,width=0.3\textwidth]{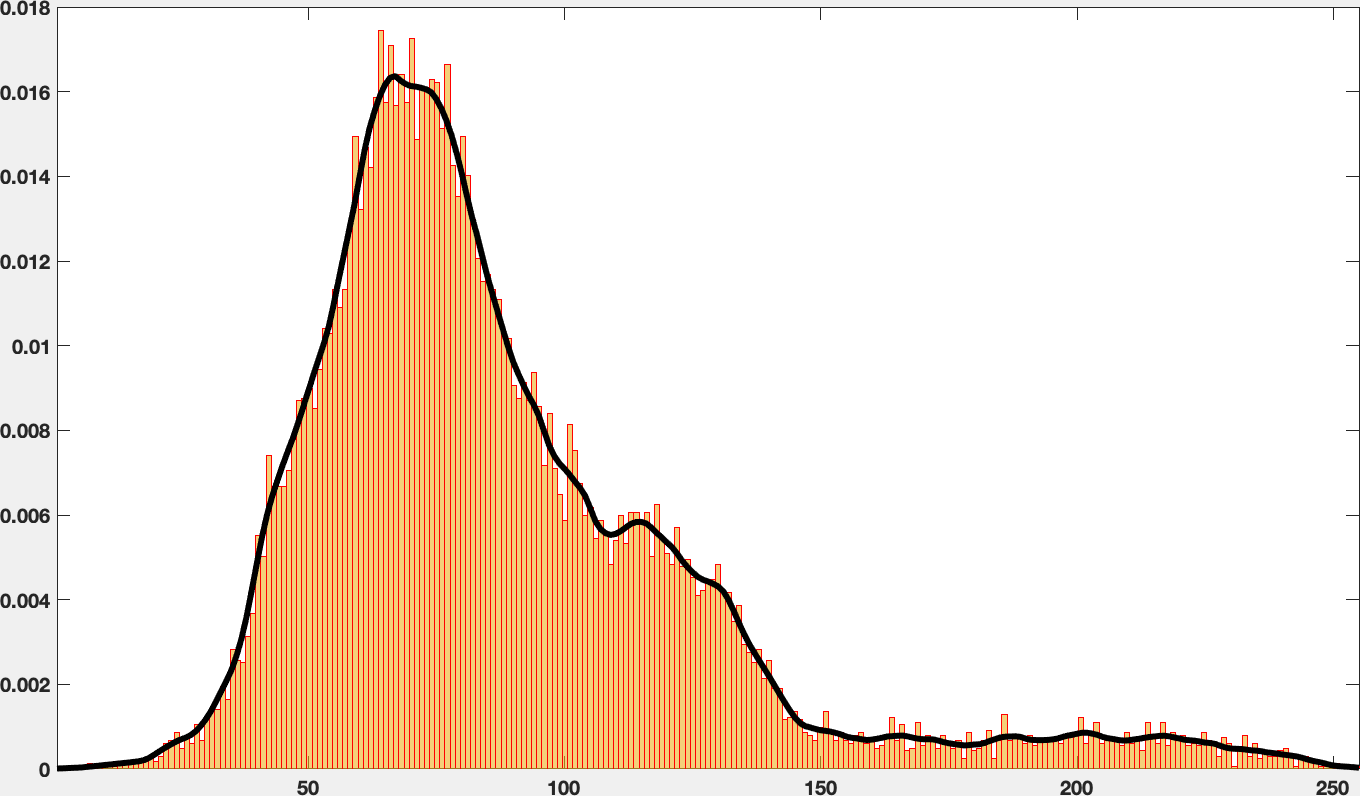}\\
         
     \end{tabular}
     \caption{Two examples from maize plants dataset where (top) is a healthy leaf and (bottom) is a leaf with disease. For each class:   an original image (left),  the extracted features (middle), and the normalized histogram (right).}
     \label{fig:ExamplesMaize}
 \end{figure}
 Fig.~\ref{fig:ExamplesMaize} illustrates an example
 of  two original images (left): a healthy plant (top) and a plant with disease (bottom), their wavelet-deconvolved versions (middle), and the corresponding histograms (right). We also display PDFs from histograms for each example in Fig.~\ref{fig:ExamplesMaize} (right column in black).
 
  A second real dataset with $1717$ observations gives the body temperature of dogs. 
	For this dataset, temporal measures of infected and uninfected dogs are stored  during $24$ hours. The infection by a parasite is suspected to cause persistent fever despite  veterinary medicine~\cite{Kumar}. The main goal is to learn the relationship between the infection and a dominant pattern from temporal temperatures. We display some examples of infected (blue) and uninfected (red) in Fig.~\ref{fig:DatasetsClassification} (c) and  we refer to this dataset as Temp.  The PDF estimates were obtained using an automatic bandwidth selection method described in~\cite{Botev-2010}. We illustrate some examples of PDFs from real datasets  in Fig.~\ref{fig:DatasetsClassification} (d\&e).
\\
We remind that high-dimensional inputs make traditional machine learning techniques  fail to solve the problem at hand. However, 
the spectral histograms as marginal distributions of the wavelet-deconvolved image  can be used to
represent/classify original images~\cite{Liu-2003}. In fact, 
instead of comparing the histograms, a better way to compare two  images (here a set of repetitive features)  would be to compare their  corresponding densities. 
 \subsubsection{Classification results}
 We learn the model parameters from $75 \%$ of
the dataset 
 whereas the rest is kept for test. This subdivision has been performed randomly $100$ times.
The performance is given as a  mean and the corresponding standard deviation (std) in order to reduce the bias (class imbalance and sample representativeness) introduced by the random  train/test split. \\
\begin{figure}[t!]
    \begin{adjustbox}{width=.9\textwidth}
        \begin{tabular}{|c|c|c|}
            \toprule
            \LARGE (\textbf{a})&
            \includegraphics[valign=m,scale=0.4]{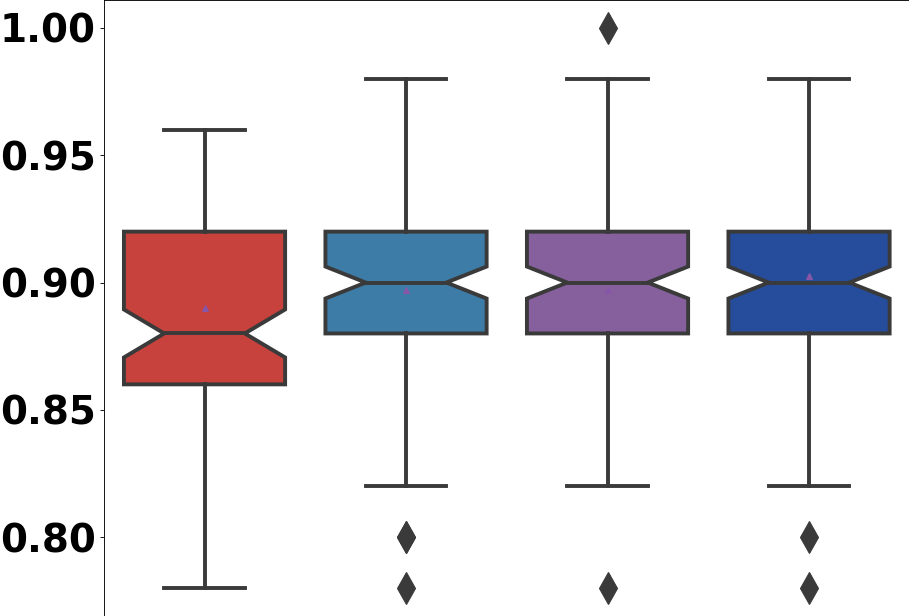} &
            \includegraphics[valign=m,scale=0.4]{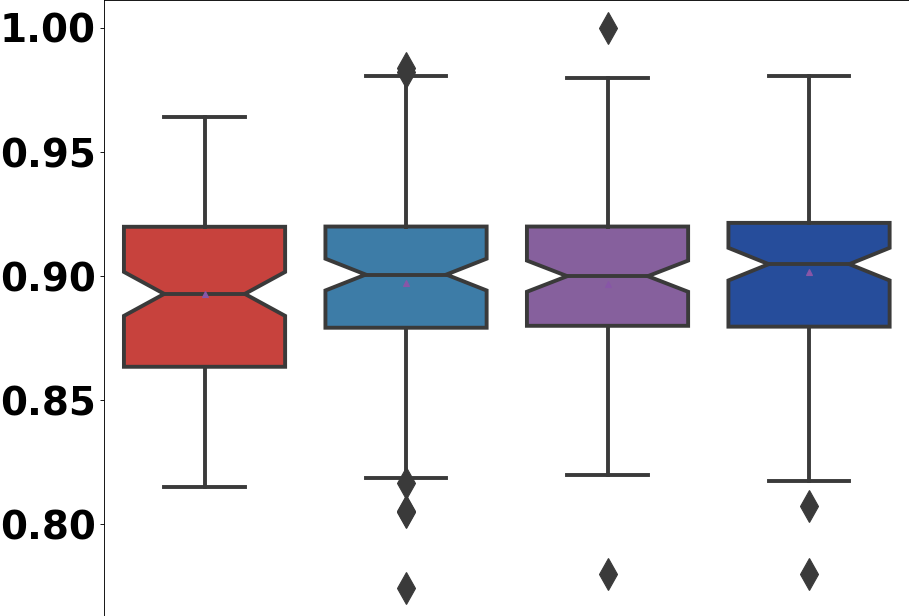}\\
            \midrule
            \LARGE (\textbf{b})&
            \includegraphics[valign=m,scale=0.4]{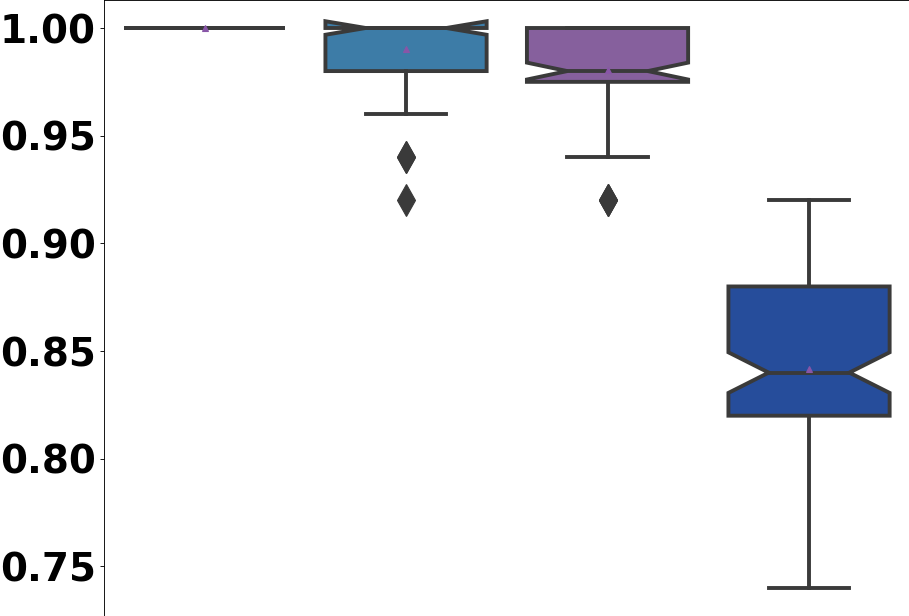}&
            \includegraphics[valign=m,scale=0.4]{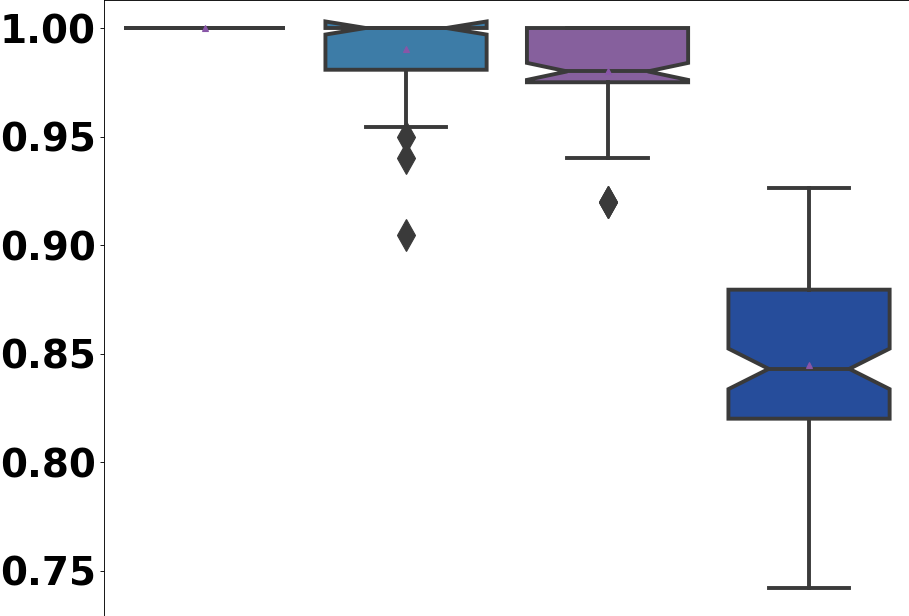}\\
            \midrule
            \LARGE (\textbf{c})&
            \includegraphics[valign=m,scale=0.4]{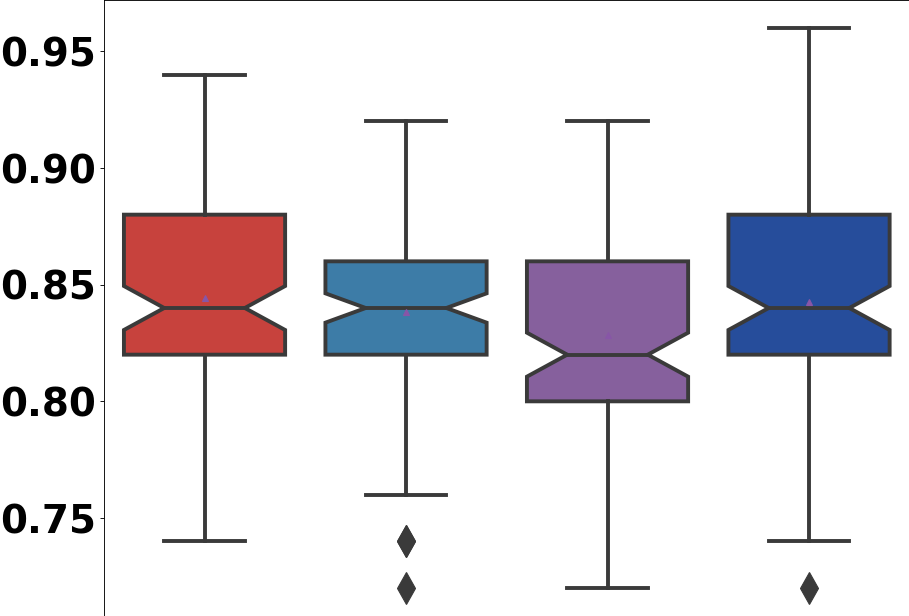}&
            \includegraphics[valign=m,scale=0.4]{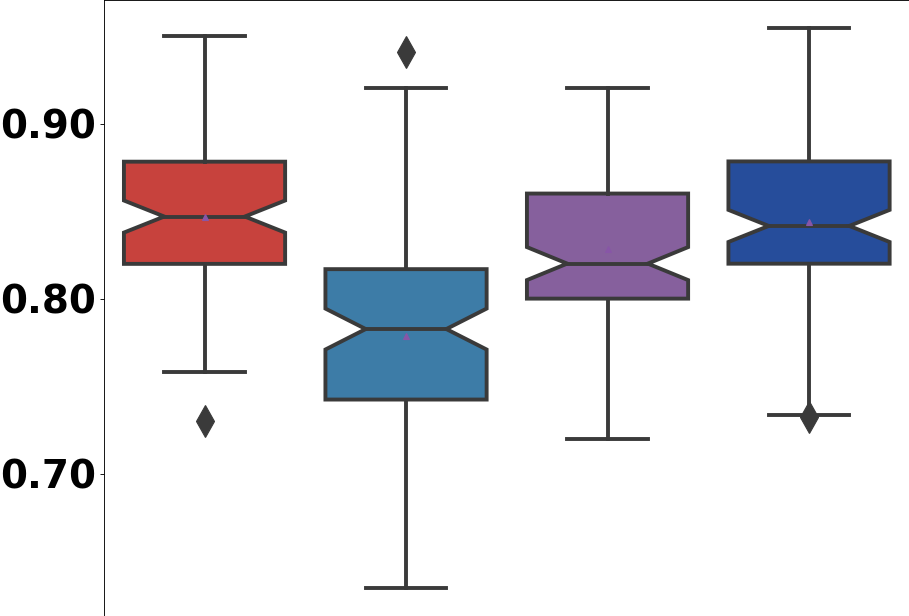}\\
            \bottomrule
        \end{tabular}
    \end{adjustbox}
    \caption{
        Boxplots of the classification accuracy (left) and AUC (right) on synthetic and semi-synthetic datasets: (a) InvGamma, (b) Beta, and (c) Growth. In each subfigure, the performance is given for different methods:
        G-GPP (red), HMC-GPP (light blue), W-GPP (violet), and JS-GPP (dark blue).}
    \label{fig:Boxplots-fig1}
\end{figure}
\begin{figure}[t!]
	\begin{adjustbox}{width=.9\textwidth}
		\begin{tabular}{|c|c|c|}
			\toprule
			\LARGE (\textbf{a})&
			\includegraphics[valign=m,scale=0.4]{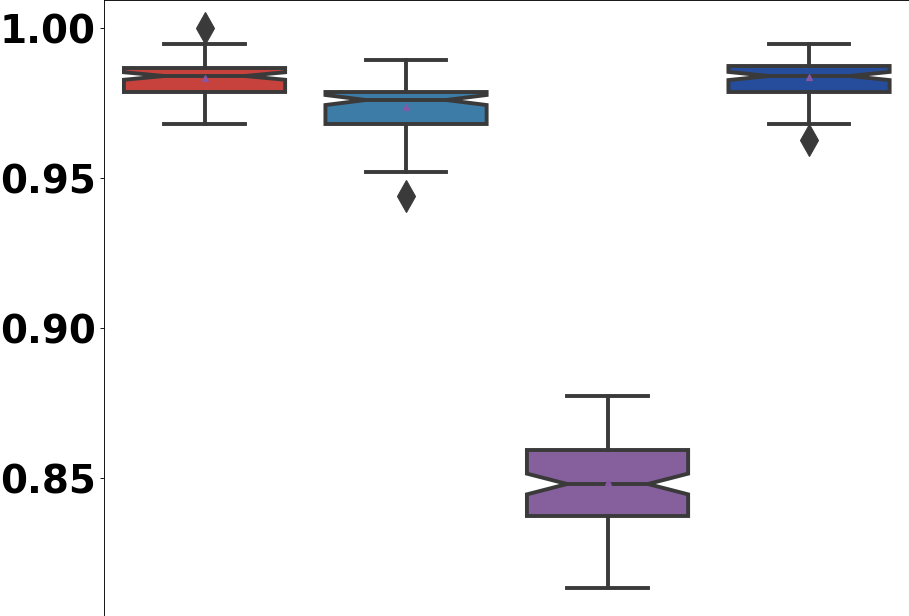}&
			\includegraphics[valign=m,scale=0.4]{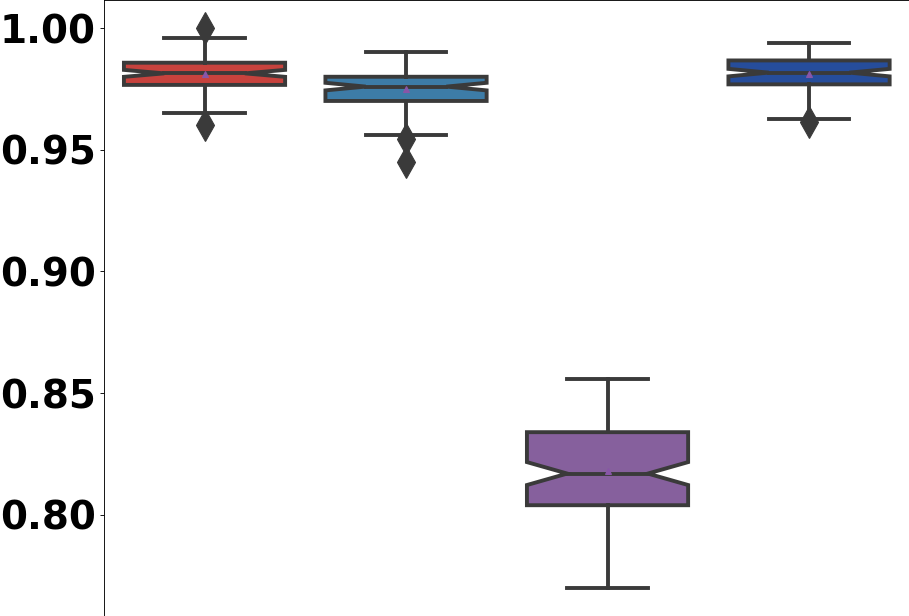}\\
			\midrule
			\LARGE (\textbf{b})&
			\includegraphics[valign=m,scale=0.4]{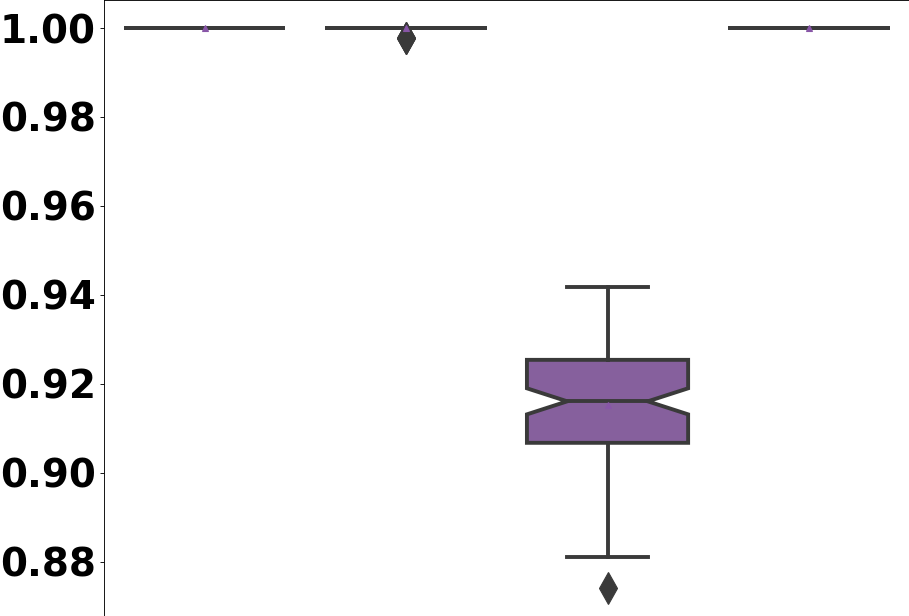}&
			\includegraphics[valign=m,scale=0.4]{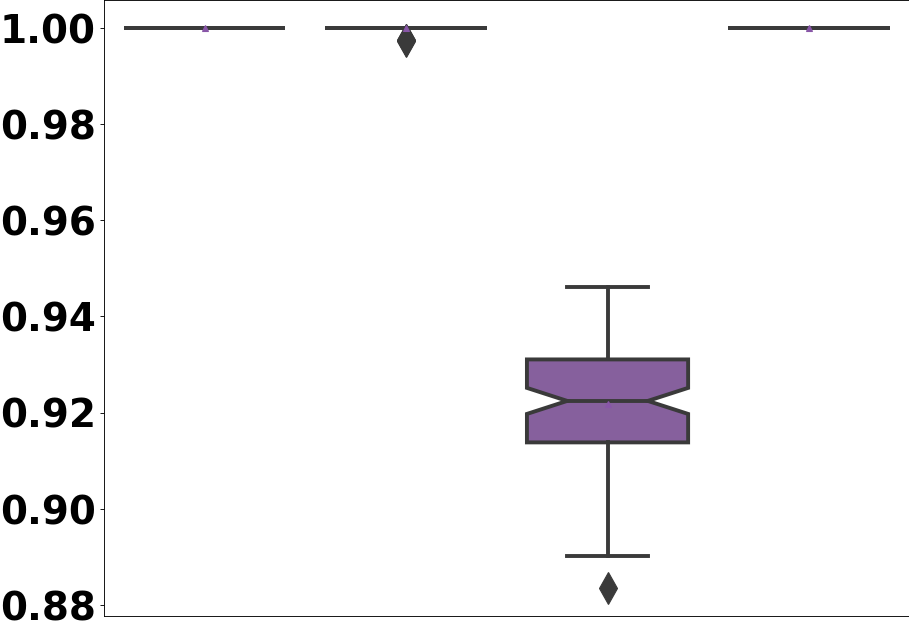}\\
			\bottomrule
		\end{tabular}
	\end{adjustbox}
	\caption{
		Boxplots of the classification accuracy (left) and AUC (right) on real datasets: (a) Temp and (b) Plants. In each subfigure, the performance is given for different methods:
		G-GPP (red), HMC-GPP (light blue), W-GPP (violet), and JS-GPP (dark blue).}
	\label{fig:Boxplots-fig2}
\end{figure}
\noindent \textbf{Results on synthetic  datasets.} We summarize all evaluation results on synthetic datasets in Fig.~\ref{fig:Boxplots-fig1} (a\&b).
Accordingly,
one can observe that both HMC-GPP, W-GPP and JS-GPP reach the best accuracy values for InvGamma with a little margin  for the proposed HMC-GPP. On the other hand, G-GPP and HMC-GPP heavily outperform W-GPP and JS-GPP for Beta.
Again, this simply shows how each  optimization method impacts  the quality of the predictive distributions.

\noindent \textbf{Results on semi-synthetic data.}
We summarize all results in Fig.~\ref{fig:Boxplots-fig1} (c) where we show  accuracy and AUC values on the Growth dataset as boxplots from $100$ tests.
 One can observe that G-GPP 
gives the best accuracy with a significant margin. Note that we have used $10^3$ HMC iterations in Algorithm~\ref{algo4}. Furthermore, 
we set the ``Burn-in" and ``Thinning" in order to ensure a fast convergence of the Markov chain and  to reduce sample autocorrelations.\\
\noindent \textbf{Results on real data.}
We further investigate whether our proposed methods can be used with  real data.
 Fig.~\ref{fig:Boxplots-fig2} (a\&b) shows the boxplots of accuracy and AUC values for Temp and  Plants, respectively. In short, 
 we highlight that the proposed methods successfully modeled these datasets with improved results in comparison with  W-GPP.
 
Fortunately, the experiments have shown that the problem of big iterations, usually needed to simulate the Markov chains for complex inputs 
 is partially solved by considering the proposed HMC sampling  (Algorithm~\ref{algo4}). In closing, we can state that the 
  leap-frog algorithm (Algorithm~\ref{algo3}),  based on Hamiltonian dynamics, allows us to early search the  best directions giving the best  minimum of the Hamiltonian defined in~(\ref{Hamiltonian}).
  
 \subsubsection{Summary of all classification results}

\begin{table}[h!]
	\centering
	\caption{Classification:  negative log-marginal likelihood as a performance metric.}
	\label{table3}
	\begin{adjustbox}{width=\textwidth}
		\begin{tabular}{|c||l|c|c|l|c|c|c|c|c|c|}
			\hline 
		\textbf{Datasets}	& \multicolumn{4}{c||}{\textbf{Synthetic}} & \multicolumn{2}{c||}{\textbf{Semi-synthetic}} &  \multicolumn{4}{c||}{\textbf{Real data}} \\
			\hline
            &\multicolumn{2}{>{\columncolor{mycolor3}}c|}{\textbf{InvGamma}}
            &\multicolumn{2}{>{\columncolor{mycolor3}}c|}{\textbf{Beta}}
            &\multicolumn{2}{>{\columncolor{mycolor1}}c|}{\textbf{Growth}}
            &\multicolumn{2}{>{\columncolor{mycolor2}}c|}{\textbf{Temp}}            
            &\multicolumn{2}{>{\columncolor{mycolor2}}c|}{\textbf{Plants}}	\\
			\cline{2-11}
			Method&mean & std& mean & std& mean & std& mean & std& mean & std
			\\
			\hline
			G-GPP & $\boldsymbol{30.50}$ & 2.43 & $\boldsymbol{4.41}$ & 0.06 & 68.03 & 3.43 & $\boldsymbol{98.66}$ & 0.73 & 98.65 & 0.72 \\
			\hline
			HMC-GPP & 105.35 &0.22 & 105.28 & 0.21& $\boldsymbol{61.65}$ &2.24 & 105.36 & 0.22 & $\boldsymbol{9.33}$ & 0.21 \\
			\hline
			JS-GPP & 32.2 &2.38 & 42.87 & 2.73& 62.0 &3.02 & 116.65 & 4.13 & 10.26 & 0.12 \\
			\hline
		\end{tabular}
	\end{adjustbox}
\end{table}

We also confirm all previous results from Table~\ref{table3}, which summarizes the mean and the std of NLML values for all datasets. 
These clearly show that at least one of the proposed methods (G-GPP or HMC-GPP) better minimizes the NLML than JS-GPP.
This brings more quite accurate estimates, which prove the predictive power of our approaches.
\section{Conclusion}
\label{sec:clc}
\noindent In this paper, we have introduced a novel framework to  extend  Bayesian learning models and   Gaussian processes when  the index support is identified with   the  space of probability density functions (PDFs). We  have detailed and applied   different numerical methods to learn regression and classification models on PDFs.  Furthermore, we  showed new theoretical results  for the Mat\'ern covariance function defined on the space of PDFs. Extensive experiments on multiple and varied  datasets have demonstrated the effectiveness and efficiency of the proposed methods in comparison with current state-of-the-art methods. 
\section*{ Acknowledgements }
\noindent This work was partially funded by the French National Centre for Scientific Research.


\begin{thebibliography}{10}

\bibitem{AbtWel1998}
M.~Abt and W.J. Welch.
\newblock Fisher information and maximum-likelihood estimation of covariance
  parameters in {G}aussian stochastic processes.
\newblock {\em The Canadian Journal of Statistics}, 26:127--137, 1998.

\bibitem{Amari-book}
S.-I. Amari.
\newblock {\em Information geometry and its applications}.
\newblock Springer, Tokyo, Japan, 1st edition, 2016.

\bibitem{Amari-Rao-87}
S.-I. Amari, O.E. Barndorff-Nielsen, R.E. Kass, S.L. Lauritzen, and C.R. Rao.
\newblock {\em Differential geometry in statistical inference}.
\newblock Institute of Mathematical Statistics, Hayward, CA, 1987.

\bibitem{Wireless-Paper}
S.~Atapattu, C.~Tellambura, and H.~Jiang.
\newblock A mixture gamma distribution to model the snr of wireless channels.
\newblock {\em IEEE Transactions on Wireless Communications}, 10:4193--4203,
  2011.

\bibitem{Atkinson-81-RaoDistance}
C.~Atkinson and A.F.S. Mitchell.
\newblock Rao's distance measure.
\newblock {\em The Indian Journal of Statistics}, 43:345--365, 1981.

\bibitem{Nihat-book}
N.~Ay, J.~Jost, H.V. Le, and L.~Schwachh\"{o}fer.
\newblock {\em Information geometry}.
\newblock Springer, Cham, Switzerland, 2017.

\bibitem{bachoc2017gaussian}
F.~Bachoc, F.~Gamboa, J-M. Loubes, and N.~Venet.
\newblock A {G}aussian process regression model for distribution inputs.
\newblock {\em IEEE Transactions on Information Theory}, 64:6620--6637, 2018.

\bibitem{Barbaresco-2013}
Fr{\'e}d{\'e}ric Barbaresco.
\newblock {\em Information Geometry of Covariance Matrix: Cartan-Siegel
  Homogeneous Bounded Domains, Mostow/Berger Fibration and Fr{\'e}chet Median},
  chapter~9, pages 199--255.
\newblock Springer, Berlin, Heidelberg, 2013.

\bibitem{Bauer-2016}
M.~Bauer, M.~Bruveris, and P.W. Michor.
\newblock Uniqueness of the {F}isher–{R}ao metric on the space of smooth
  densities.
\newblock {\em Bulletin of the London Mathematical Society}, 48:499--506, 2016.

\bibitem{bauer-metrics}
M.~Bauer, E.~Klassen, S.C. Preston, and Z.~Su.
\newblock A diffeomorphism-invariant metric on the space of vector-valued
  one-forms, 2018.

\bibitem{Paper-Bhattacharyya43}
A.~Bhattacharyya.
\newblock On a measure of divergence between two statistical populations
  defined by their probability distributions.
\newblock {\em Bulletin of the Calcutta Mathematical Society}, 35:99--109,
  1943.

\bibitem{Botev-2010}
Z.I. Botev, J.F. Grotowski, and D.P. Kroese.
\newblock Kernel density estimation via diffusion.
\newblock {\em The Annals of Statistics}, 38:2916--2957, 2010.

\bibitem{Cencov-book-1982}
N.N. Cencov.
\newblock {\em Statistical decision rules and optimal inference}.
\newblock Translations of Mathematical Monographs. American Mathematical
  Society, Providence, R.I., 1982.

\bibitem{DeChant}
C.~DeChant, T.~Wiesner-Hanks, S.~Chen, E.~Stewart, J.~Yosinski, M.~Gore,
  R.~Nelson, and H.~Lipson.
\newblock Automated identification of northern leaf blight-infected {M}aize
  plants from field imagery using deep learning.
\newblock {\em Phytopathology}, 107:1426--1432, 2017.

\bibitem{Duane1987216}
S.~Duane, A.~Kennedy, B.~Pendleton, and D.~Roweth.
\newblock Hybrid {M}onte {C}arlo.
\newblock {\em Physics Letters B}, 195:216--222, 1987.

\bibitem{Friedrich91}
T.~Friedrich.
\newblock Die {F}isher-information und symplektische strukturen.
\newblock {\em Mathematische Nachrichten}, 153:273--296, 1991.

\bibitem{Gelman06priordistributions}
A.~Gelman.
\newblock Prior distributions for variance parameters in hierarchical models.
\newblock {\em Bayesian Analysis}, 1:515--533, 2006.

\bibitem{genton2015}
M.G. Genton and W.~Kleiber.
\newblock Cross-covariance functions for multivariate geostatistics.
\newblock {\em Statistical Science}, 30:147--163, 2015.

\bibitem{Helgason1978}
S.~Helgason.
\newblock {\em Differential geometry, lie groups, and symmetric spaces}.
\newblock Academic Press, New York, 1978.

\bibitem{NIPS2011_4241}
D.~Hern\'{a}ndez-Lobato, J.M. Hern\'{a}ndez-lobato, and P.~Dupont.
\newblock Robust multi-class {G}aussian process classification.
\newblock In {\em Proceedings of the 24th International Conference on Neural
  Information Processing Systems}, NIPS'11, pages 280--288, Red Hook, NY, USA,
  2011. Curran Associates, Inc.

\bibitem{Mitsuhiro-Geodesic-2015}
M.~Itoh and H.~Satoh.
\newblock Geometry of {F}isher information metric and the barycenter map.
\newblock {\em Entropy}, 17:1814--1849, 2015.

\bibitem{Kuczmarski}
R.J. Kuczmarski, C.~Ogden, S.S. Guo, L.~Grummer-Strawn, K.M. Flegal, Z.~Mei,
  R.~Wei, L.R. Curtin, A.F. Roche, and C.L. Johnson.
\newblock 2000 {CDC} growth charts for the united states: methods and
  development.
\newblock {\em Vital and Health Statistics}, 246:1--190, 2002.

\bibitem{Kumar}
P.~Kumar and A.~Kumar.
\newblock Haemato-biochemical changes in dogs infected with {B}abesiosis.
\newblock In {\em Conference on Food Security and Sustainable Agriculture},
  pages 21--24. International Journal of Chemical Studies, 2018.

\bibitem{Hyperbolic-Lee}
J.M. Lee.
\newblock {\em Riemannian manifolds: An introduction to curvature}.
\newblock Springer Science, New York, 1997.

\bibitem{Nielsen2013}
F.~Nielsen and~M. Liu and B.C. Vemuri.
\newblock {\em Jensen divergence-based means of {SPD} matrices}, chapter~6,
  pages 111--122.
\newblock Springer, Berlin, Heidelberg, 2013.

\bibitem{Liu-2003}
X.~Liu and D.~Wang.
\newblock Texture classification using spectral histograms.
\newblock {\em IEEE Transactions on Image Processing}, 12:661--670, 2003.

\bibitem{NIPS2017_7149}
A.~Mallasto and A.~Feragen.
\newblock Learning from uncertain curves: the 2-{W}asserstein metric for
  {G}aussian processes.
\newblock In {\em Neural Information Processing Systems (NIPS)}, pages
  5660--5670. Curran Associates, Inc., Long Beach, CA, USA, 2017.

\bibitem{Neal97montecarlo}
R.M. Neal.
\newblock {\em Monte {C}arlo implementation of {G}aussian process models for
  {B}ayesian regression and classification}.
\newblock Technical report. University of Toronto (Dept. of Statistics),
  Toronto, Canada, 1997.

\bibitem{1206.1901}
R.M. Neal.
\newblock {MCMC} using {H}amiltonian dynamics.
\newblock {\em Handbook of {M}arkov {C}hain {M}onte {C}arlo}, 54:113--162,
  2010.

\bibitem{Nguyen-Div-JS}
H.V. Nguyen and J.~Vreeken.
\newblock Non-parametric {J}ensen-{S}hannon divergence.
\newblock In {\em Machine Learning and Knowledge Discovery in Databases}, pages
  173--189, Cham, Switzerland, 2015. Springer International Publishing.

\bibitem{DBLP:conf/aistats/OlivaNPSX14}
J.B. Oliva, W.~Neiswanger, B.~P{\'{o}}czos, J.~G. Schneider, and E.~P. Xing.
\newblock Fast distribution to real regression.
\newblock In {\em Proceedings of the Seventeenth International Conference on
  Artificial Intelligence and Statistics}, Proceedings of Machine Learning
  Research, pages 706--714, Reykjavik, Iceland, 2014. PMLR.

\bibitem{Pistone-95}
G.~Pistone and C.~Sempi.
\newblock An infinite-dimensional geometric structure on the space of all the
  probability measures equivalent to a given one.
\newblock {\em The Annals of Statistics}, 23:1543--1561, 1995.

\bibitem{pmlr-v31-poczos13a}
B.~P{\'{o}}czos, A.~Singh, A.~Rinaldo, and L.~Wasserman.
\newblock Distribution-free distribution regression.
\newblock In {\em Proceedings of the Sixteenth International Conference on
  Artificial Intelligence and Statistics}, pages 507--515, Scottsdale, Arizona,
  USA, 2013. PMLR.

\bibitem{Ramsay-1991}
J.O. Ramsay and B.W. Silverman.
\newblock {\em Functional Data Analysis}.
\newblock Springer-Verlag, New York, USA, 2005.

\bibitem{Rao-45}
C.R. Rao.
\newblock Information and the accuracy attainable in the estimation of
  statistical parameters.
\newblock {\em Bulletin of Calcutta Mathematical Society}, 37:81--91, 1945.

\bibitem{Rao-82}
C.R. Rao.
\newblock Diversity and dissimilarity coefficients: A unified approach.
\newblock {\em Theoretical Population Biology}, 21:24--43, 1982.

\bibitem{rasmussen06gaussian}
C.E. Rasmussen and C.K.I. Williams.
\newblock {\em Gaussian processes for machine learning}.
\newblock The MIT Press, Cambridge, London, 2006.

\bibitem{Samir-FoCom12}
C.~Samir, P.-A. Absil, A.~Srivastava, and E.~Klassen.
\newblock A gradient-descent method for curve fitting on {R}iemannian
  manifolds.
\newblock {\em Foundations of Computational Mathematics}, 12:49--73, 2012.

\bibitem{Samir-wacv-16}
C.~Samir, S.~Kurtek, A.~Srivastava, and N.~Borges.
\newblock An elastic functional data analysis framework for preoperative
  evaluation of patients with rheumatoid arthritis.
\newblock In {\em Winter Conference on Applications of Computer Vision}, pages
  1--8, Lake Placid, NY, USA, 2016. {IEEE}.

\bibitem{shishido2005}
Y.~Shishido.
\newblock Strong symplectic structures on spaces of probability measures with
  positive density function.
\newblock {\em Proceedings of the Japan Academy, Series A, Mathematical
  Sciences}, 81:134--136, 2005.

\bibitem{Fukumizu-2013}
B.~Sriperumbudur, K.~Fukumizu, A.~Gretton, A.~Hyv{\"a}rinen, and R.~Kumar.
\newblock Density estimation in infinite dimensional exponential families,
  2013.

\bibitem{5513626}
B.K. Sriperumbudur, K.~Fukumizu, A.~Gretton, B.~Sch\"{o}lkopf, and G.R.G.
  Lanckriet.
\newblock Non-parametric estimation of integral probability metrics.
\newblock In {\em International Symposium on Information Theory (ISIT)}, pages
  1428--1432, Piscataway, NJ, USA, 2010. IEEE.

\bibitem{Srivastava-2007}
A.~Srivastava, I.~Jermyn, and S.~Joshi.
\newblock Riemannian analysis of probability density functions with
  applications in vision.
\newblock In {\em Conference on Computer Vision and Pattern Recognition
  (CVPR)}, pages 1--8, Minneapolis, USA, 2007. IEEE.

\bibitem{Anuj-book-2016}
A.~Srivastava and E.~Klassen.
\newblock {\em Functional and shape data analysis}.
\newblock Springer-Verlag, New York, USA, 2016.

\bibitem{Ste1999}
M.L. Stein.
\newblock {\em Interpolation of spatial data}.
\newblock Springer-Verlag, New York, USA, 1999.

\bibitem{DBLP:journals/corr/SutherlandOPS15}
D.J. Sutherland, J.B. Oliva, B.~P{\'{o}}czos, and J.G. Schneider.
\newblock Linear-time learning on distributions with approximate kernel
  embeddings.
\newblock In {\em Proceedings of the Thirtieth AAAI Conference on Artificial
  Intelligence}, AAAI'16, pages 2073--2079, Phoenix, Arizona, 2016. AAAI Press.

\bibitem{Zhang-2019}
Z.~Zhang, E.~Klassen, and A.~Srivastava.
\newblock Robust comparison of kernel densities on spherical domains.
\newblock {\em Sankhya A: The Indian Journal of Statistics}, 81:144--171, 2019.

\end{thebibliography}
\end{document}